\documentclass[preprint2,usenatbib]{mnras}
\usepackage{graphicx}
\usepackage {amsmath,amssymb}
\usepackage{color}
\usepackage{aas_macros}
\usepackage{array}
\usepackage{booktabs}

\newcommand{\Msun}{{\rm M}_{\odot}}

\usepackage[dvipsnames]{xcolor}

\title{Environmental effects on Low Surface Brightness Galaxies in the IllustrisTNG simulation}

\author[P\'erez-Monta\~no, et al.]
{
	\parbox{18cm}{
		Luis Enrique P\'erez-Monta\~no,$^{1,2}$\thanks{E-mail: le.perezmontano@zju.edu.cn}
        Bernardo Cervantes Sodi,$^{2}$\thanks{E-mail: b.cervantes@irya.unam.mx}
        Vicente Rodriguez-Gomez,$^{2}$
        Qirong Zhu$^{3}$ and Go Ogiya$^{1}$.
	}
	\vspace{0.3cm} \\ 
    $^{1}$ Institute for Astronomy, School of Physics, Zhejiang University, Hangzhou 310027, People’s Republic of China\\
	$^{2}$ Instituto de Radioastronomía y Astrofísica, Universidad Nacional Autónoma de México, C.P. 58089, Morelia, Michoacán, México \\
	$^{3}$ McWilliams Center for Cosmology, Department of Physics, Carnegie Mellon University, Pittsburgh, PA 15213, USA \\
}

\date{Accepted XXX. Received YYY; in original form ZZZ}

\pubyear{2024}

\begin{document}
\label{firstpage}
\pagerange{\pageref{firstpage}--\pageref{lastpage}}
\maketitle

\begin{abstract}
Employing the TNG100 run of the IllustrisTNG project, we characterize the environment of Low Surface Brightness Galaxies (LSBGs) across varying scales, from their associated dark matter halos to their distribution within the broader cosmic structure. We find no significant differences in the halo concentration index $c_{200}$ between LSBGs and their High Surface Brightness (HSBGs) counterparts, with LSBGs residing in halos with higher spin parameter $\lambda$ and slightly more spherical shapes than HSBGs. LSBGs show a stronger alignment between the dark and stellar angular momentum vectors than their high surface brightness counterparts. The relative abundance of LSBGs within groups and clusters displays a central deficit, hinting at potential destruction upon reaching these core regions. Studying the density field, we find a preference for rotation-dominated LSBGs to reside in low-density environments, while dispersion-dominated LSBGs thrive in high-density regions where galaxy interactions govern their evolution, an observation corroborated by our analysis of the two-point correlation function $\xi (r)$. Our examination of the cosmic web reveals no significant differences in the distance to the closest large-scale structure, barring a few exceptions. This suggests a limited impact of large-scale spatial distribution on mechanisms driving LSBG evolution. All together, we conclude that the halo vicinity and local environment at the scale of galaxy clusters, where mechanisms such as galaxy mergers and tidal stripping, as well as stellar and gas accretion take place, is the most likely environment that favour the emergence of LSBGs with different morphologies, mostly driven by the presence or absence of important local interaction phenomena.

\end{abstract}




\section{Introduction}
\label{sec:Intro}
	Low Surface Brightness galaxies (LSBGs) are objects characterized to be optically fainter \citep{Impey97}, bluer and having lower stellar densities \citep{Vorobyov09} than their High Surface Brightness counterparts (HSBGs). Typically, LSBGs are defined as galaxies with a central surface brightness $\mu_B \gtrsim $ 21.65 mag arcsec$^{-2}$ \citep{Freeman70}. However, it is also possible to classify them using different wavelengths, as many authors do. For instance, \citet{Zhong08} and \citet{Bakos12} employ the $ r-$band of the SDSS to more accurately map the underlying stellar mass, rather than recent formation processes. While these objects were not commonly found in the past, the development of increasingly sophisticated instruments has led to the discovery of a growing number of LSBGs. As of today, they are thought to constitute a significant fraction of extragalactic sources, accounting for, in number, up to $\sim$ 30 - 50\% of the population of extragalactic source population \citep{McGaugh95a, ONeil03}. The increasing discovery of LSBGs in the local universe strongly suggests that they constitute a substantial portion of the baryonic budget of the Universe.
		
	Although LSBGs exhibit various morphologies, they are predominantly characterized as late-type galaxies, lacking prominent bulges (with bulge-to-disk luminosity ratios below 0.1,\citealt{McGaugh95b}), as well as strong bars \citep{Honey16, Cerv17}. Additionally, LSBGs tend to be quiescent \citep{Wyder09,Schombert11}, gas-rich \citep{Huang14,Du15}, metal-poor ($Z < $ 0.003, \citealt{deBlok98a,deBlok98b,deNaray04}) and dust-free \citep{Hinz07,Rahman07}. These characteristics pose challenges for the formation of giant molecular clouds and the initiation of intense star-forming episodes. Notably, a minority of LSBGs harbor an Active Galactic Nucleus (AGN, \citealt{Impey96, Galaz11}), typically hosting less massive supermassive black holes (SMBHs) with lower accretion rates than HSBGs of similar stellar mass \citep{Subramanian16,Saburova21}. 
 
	LSBGs have garnered considerable attention from researches owing to their intriguing characteristics. It has been found that these galaxies are mostly dominated by dark matter (DM) at all radii \citep{deBlok01}, and several studies suggest that LSBGs reside within DM halos with high angular momentum, acquisition attributed to tidal forces experienced by the primordial proto-galaxies exerted by neighbouring overdensities in the early Universe \citep{Hoyle51,Peebles69}. Within this framework, in cases where the baryonic-to-halo specific angular momentum ratio approximates unity, certain structural characteristics become contingent upon the spin of the DM halo. This includes parameters such as the scale-length of the disks and their stellar surface density \citep{Fall80,Hdz98,Mo98,HdzCerv06,Cervantes13}. The spin parameter \citep{Peebles69,Peebles71} serves as a dimensionless metric quantifying the rotational support within galaxies, effectively denoting the ratio of centripetal to gravitational acceleration.The large amount of angular momentum, combined with high values of their spin parameter are thought to be the main reason of their low stellar densities and therefore, their surface brightness \citep{Dalcanton97,Boissier03,KimLee13,PerezMontano19}.
	
	While numerous studies have probed the intrinsic properties of LSBGs through a combination of observational data and simulations, the influence of their environment on formation and evolution remains a subject of ongoing ambiguity. Previous studies regarding the environment of LSBGs have predominantly centered on either small (hundreds of kpc) or large-scale ranges (tens of Mpc), primarily due to inherent technical constrains on observational studies. Interestingly, despite such limitations and the different scales explored by different authors, a recurrent finding across these studies is the relative isolation in which LSBGs are typically found.

\begin{figure*} 
\begin{tabular}{cccc}
\centering
     \includegraphics[width=0.23\textwidth]{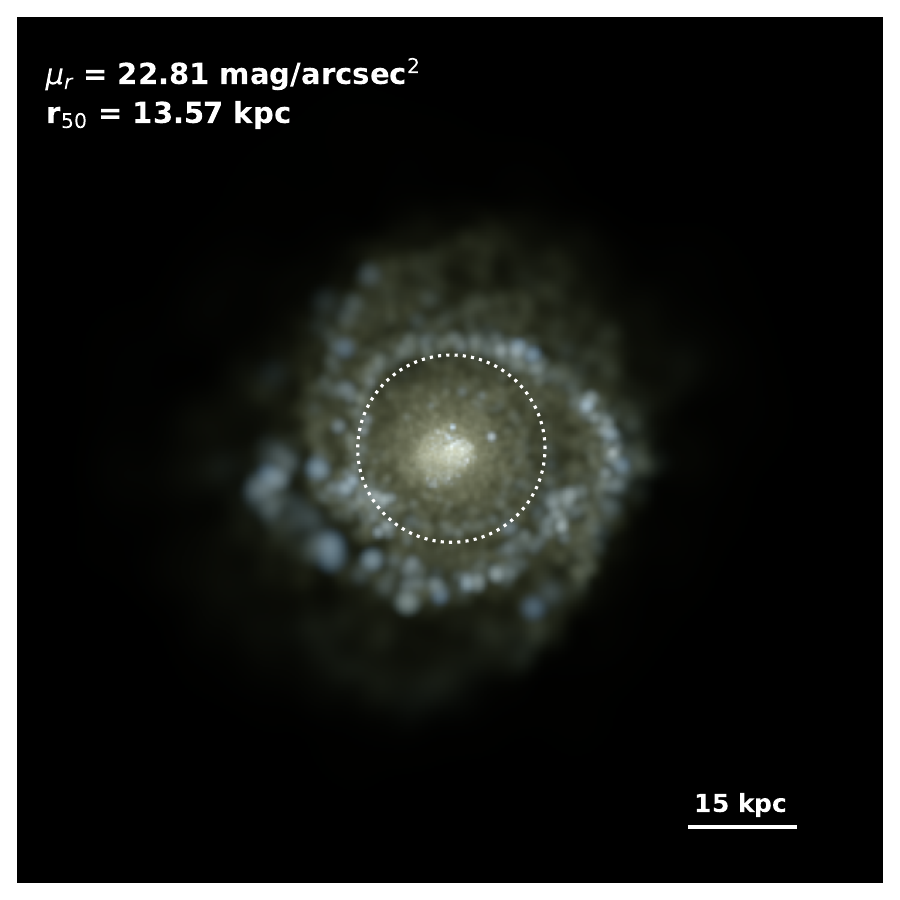} &  \includegraphics[width=0.23\textwidth]{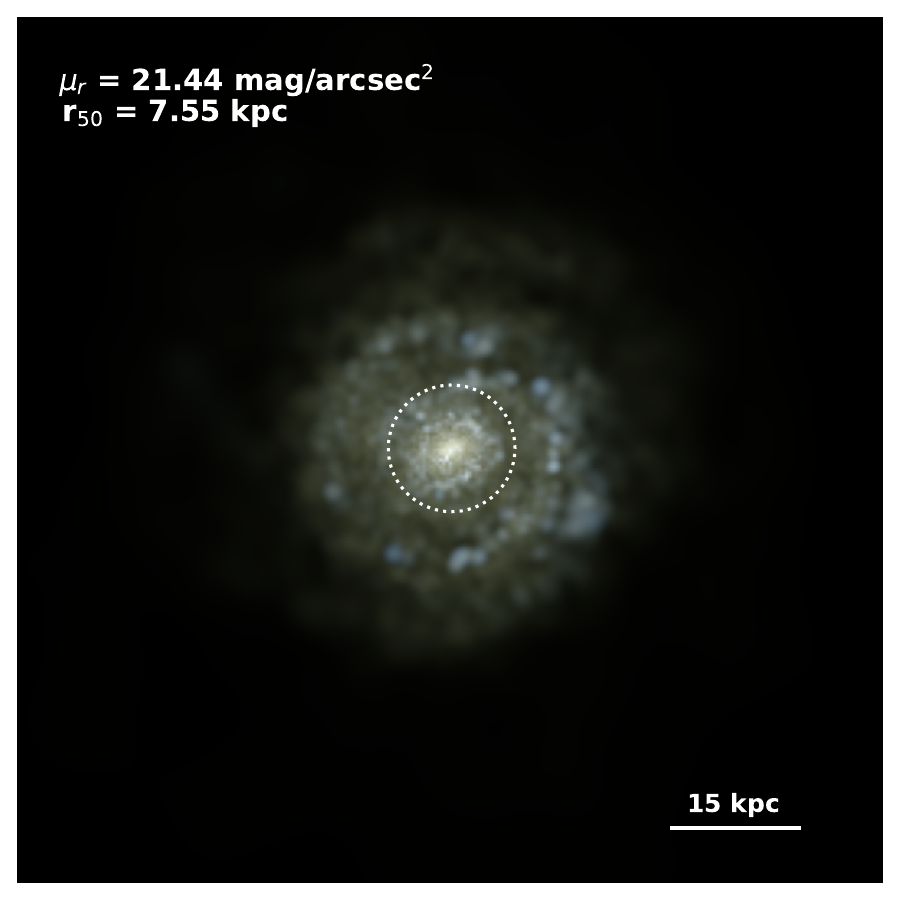} &
     \includegraphics[width=0.23\textwidth]{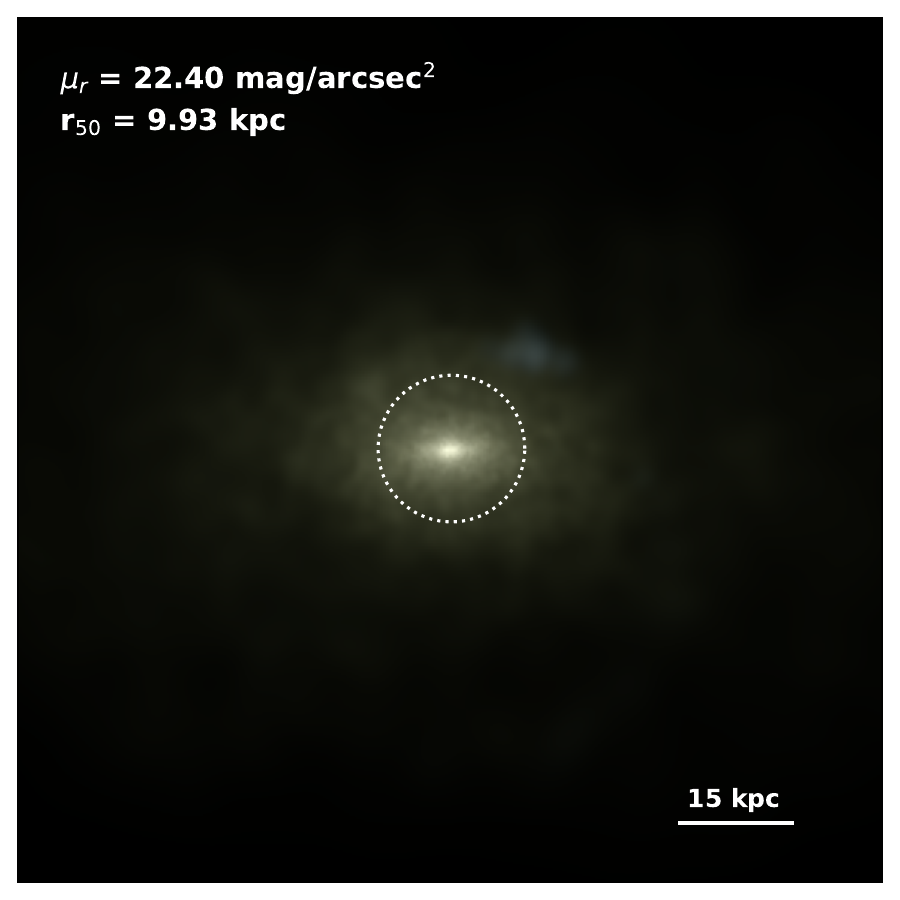} &
     \includegraphics[width=0.23\textwidth]{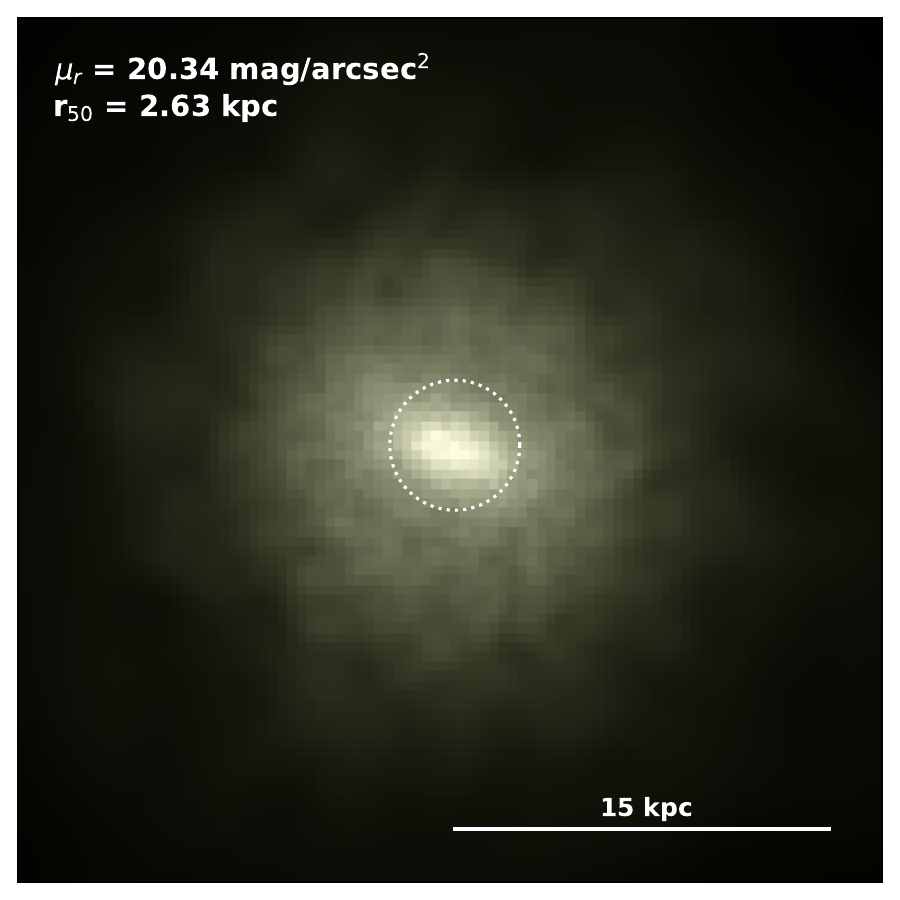} \\ \includegraphics[width=0.23\textwidth]{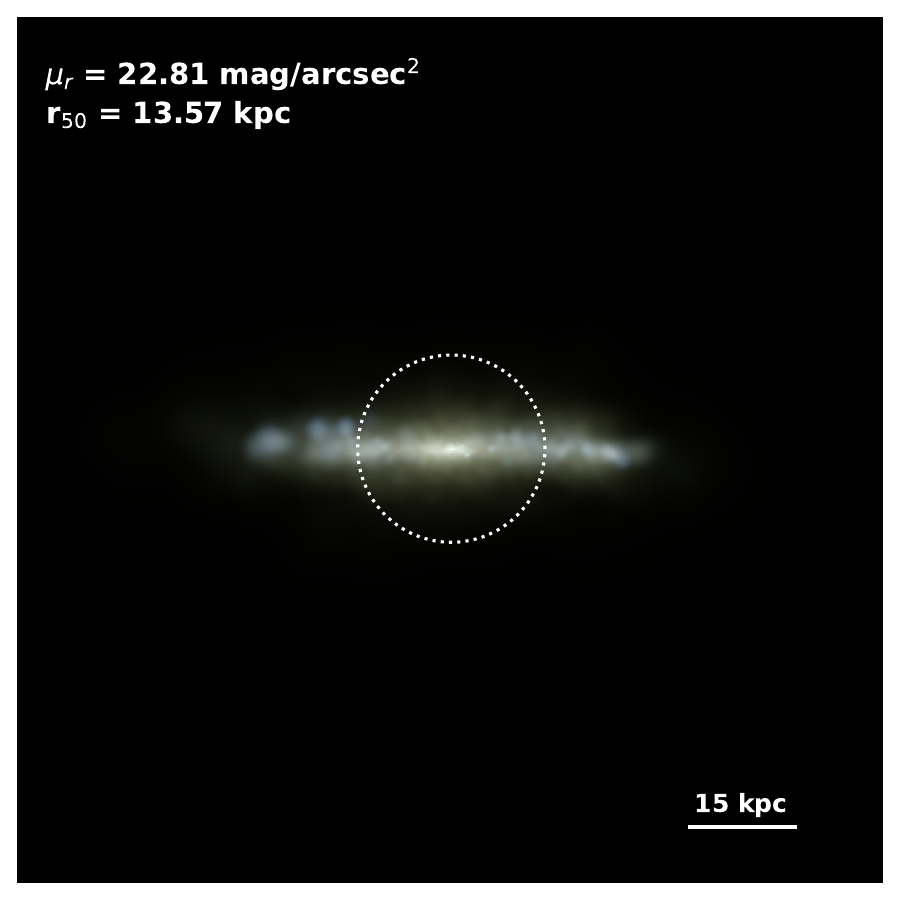} &  \includegraphics[width=0.23\textwidth]{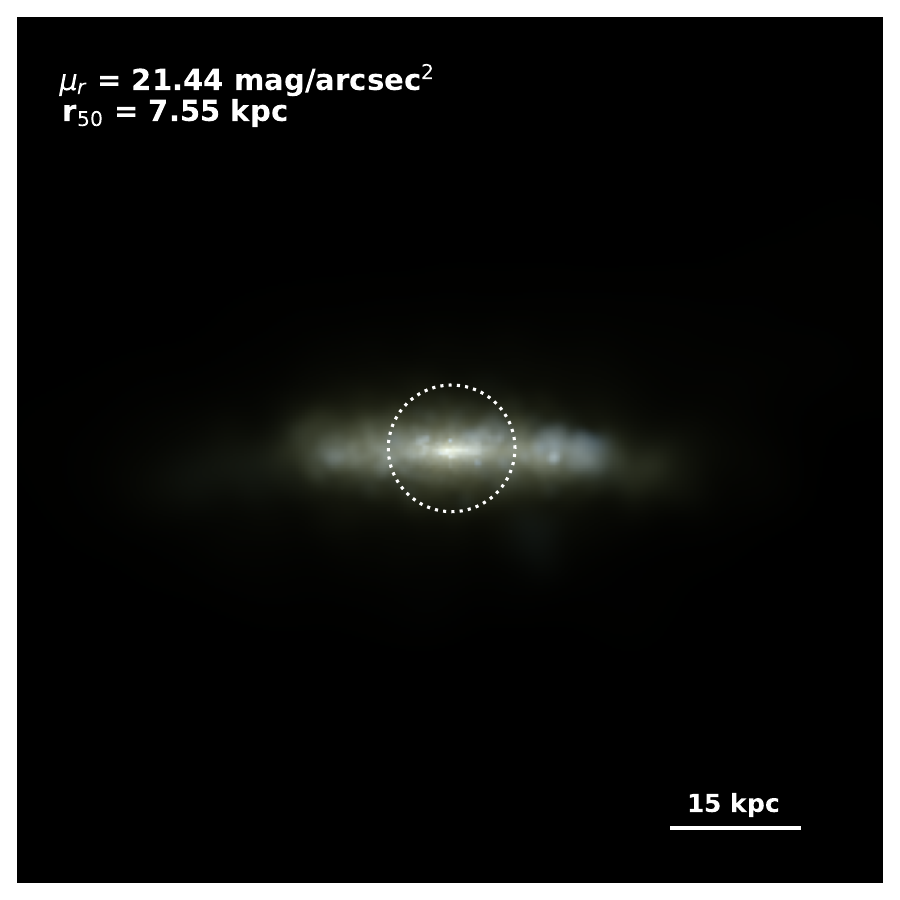} &
     \includegraphics[width=0.23\textwidth]{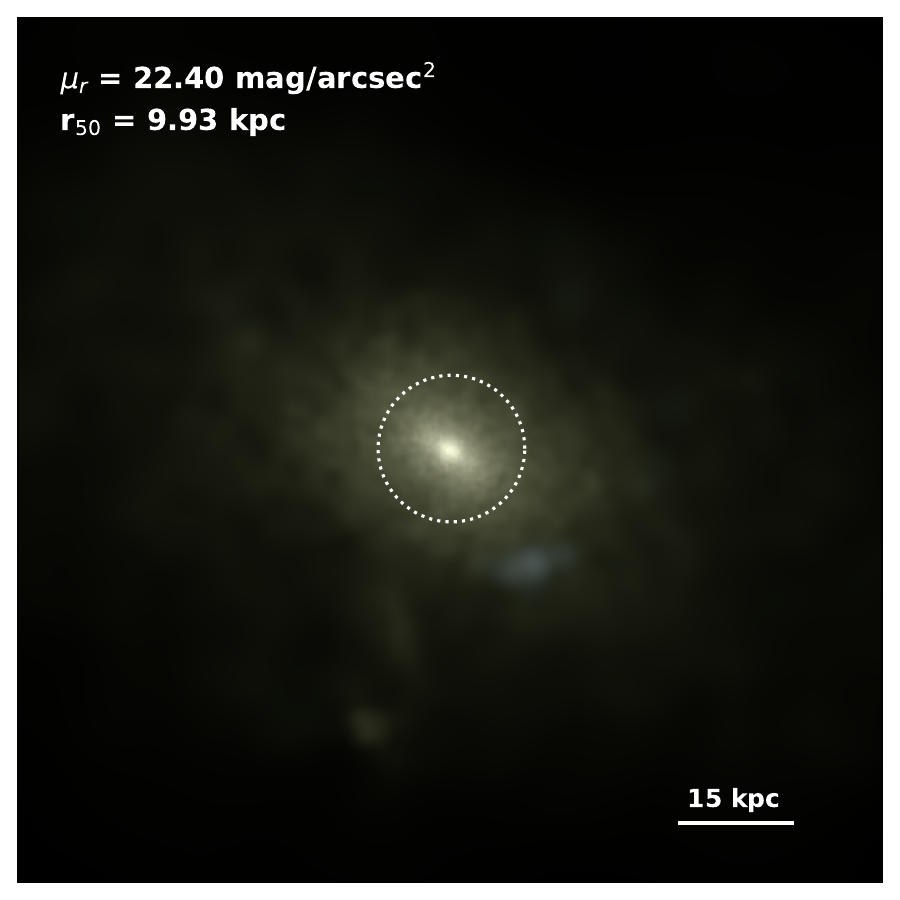} &
     \includegraphics[width=0.23\textwidth]{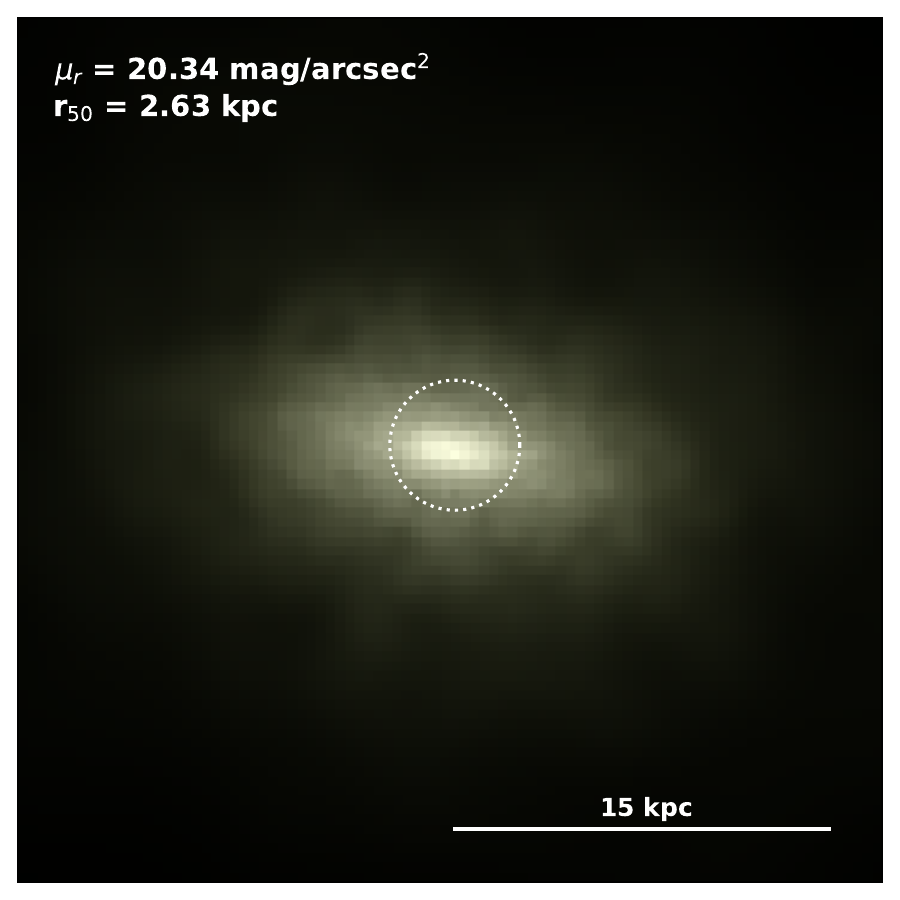} \\
\end{tabular}
\caption{Synthetic images of four simulated galaxies with $M_{*} \sim 10^{10.5} \Msun$ and different surface brightness, including rotation-dominated (first and second columns) and dispersion-dominated (third and fourth columns) galaxies. The top row includes the face-on view of the randomly selected galaxies, while the bottom row corresponds to the edge-on view of them.}
\label{fig:images}
\end{figure*}

	In \cite{Bothun93}, the authors found that the average distance between LSBGs and their closest neighbour was about 1.7 times larger than that of their HSBGs counterparts. \citet{Rosenbaum09} employed a sample of galaxies at $z \leq 0.1$ drawn from the SDSS DR4, finding that these galaxies are found in more isolated environments than HSBGs within scales of 2-5 Mpc. Similarly, \citet{Galaz11} reported a similar lack of neighbours for LSBGs (about 76\% and 70\% for LSBGs and HSBGs, respectively) at $r < 0.5$ Mpc. This particularity has a huge impact on galaxy evolution, suggesting that the relatively isolation of these kind of galaxies is an important condition for their survival due to the difficulty to amplify and propagate perturbations in such conditions. In \citet{PerezMontano19} it is found that the fraction of isolated central LSBGs is higher than the fraction of isolated HSBGs. 

    Interestingly, these previous results seem to be in contrast with \citet{Wang11}, where a mild but significant correlation between the spin parameter and the environment in which galaxies reside is found. These authors employed a set of high-resolution N-body simulations, finding that the spin parameter increases when the tidal field is more intense, as well as with the local overdensity, implying that stronger gravitational interactions are reflected in high spin halos (as are the case of LSBGs). 
	
	Very few works have made emphasis on the study of the large-scale environment of LSBGs, especially from the observational point of view. \citet{Rosenbaum09} suggested that LSBGs are formed originally inside low-density structures and then, at later epochs, these are displaced towards the external parts of filaments and walls of the cosmic web. \citet{Ceccarelli12} found a strong systematic drop in the LSBG fraction when moving away from the center of voids, while the fraction of HSBGs is increased. Moreover, the transformation of quiescent LSBGs into star-forming HSBGs is possibly due to gas arriving from the void interior as a consequence of void expansion. Finally \citet{PerezMontano19} found a slight preference of LSBGs to reside in filaments instead of clusters, and the distance to the nearest filament being larger when compared with HSB ones, however, this difference is relatively small when compared to other properties at smaller scales, such as the spin parameter of the dark matter halo in which LSBGs reside. This result led to the conclusion that the main differences in the environment around LSBGs and HSBGs lie on small scales, rather than the large-scale ones. 

	The use of hydrodynamical cosmological simulations has become an  important tool towards the understanding of the mechanisms for the formation and evolution of LSBGs \citep{Zhu18,DiCintio19,PerezMontano22}. Particularly, using simulated galaxies from EAGLE \citep{Schaye15,Crain15,McAlpine16}, \citet{Kulier20} found that LSBGs are further away from their closest neighbour than HSBGs, implying that the former are generally more isolated not because they formed in low-density environments, but due to the destruction of their closest neighbors which eventually leds to the formation of extended disks around LSBGs. On the other hand, \citet{Martin19} employed a sample of simulated galaxies drawn from Horizon AGN (\citealt{Dubois14}) in order to study the spatial distribution of LSBGs, finding that these galaxies are found in high density environments when compared with HSBGs, but LSBGs having lower spin parameters than HSBGs, which results at odds with the results by \citet{Wang11}.
 
    Clearly these previous studies show contrasting results, often lacking compatibility with one another. Consequently, there is a pressing need to investigate the environmental influence through diverse quantification methods across varying scales. For that purpose, in this work we employ a sample of simulated galaxies derived from IllustrisTNG to study the environment of LSBGs at different scales. This approach aims to a more comprehensive and nuanced examination of the environmental factors impacting LSBGs, shedding light on the underlying causes of these disparities, as well as a better understanding of the main phenomena that play an important role in determining the evolutionary path that these galaxies follow, culminating in their emergence as LSBGs. The paper is organized as follows: In Section \ref{sec:Sample}, we describe the simulated sample employed in this work. In Section \ref{sec:Results}, we present the main results of our work, describing the different characterization of the environment at different scales, going from the smallest scales comparable with galaxy halos, LSBGs within galaxy groups, and the characterization of the large-scale structure of the Universe, and finally in Section \ref{sec:Conclusions}, we provide a general discussion and a brief summary of our results.

\section{Simulated Sample}
\label{sec:Sample}

\subsection{The TNG simulation}
The IllustrisTNG project (hereafter TNG, \citealt{Nelson18,Pillepich18,Springel18}) constitute an improved version of the original Illustris galaxy formation model \citep{Vogelsberger13,Genel14,Torrey14,Vogelsberger14,Nelson15}, which includes 18 different magneto-hydrodynamical cosmological simulations, each one including different box sizes, resolution and physical processes. Using the moving-mesh code \texttt{AREPO} \citep{Arepo10}, TNG follows the gas dynamics coupled with DM using a quasi-Lagrangian treatment, in which a dynamical adaptive discretization allows the solution of the hydrodynamical equations to move together with the gas. DM halos in the simulation are identified employing a Friends-of-Friends algorithm (FoF, \citealt{Davis85}). The \texttt{SUBFIND} algorithm \citep{Dolag09,Springel01} is employed to identify gravitationally bound substructures enclosed by each FoF halo, which represent the presence of subhalos within them.

The simulation includes boxes of approximately 50, 100 and 300 Mpc per side, namely TNG50, TNG100 and TNG300 respectively, and each of them consist on 100 snapshots running from $z=20$ to $z=0$. Additionally, every box runs three simulations at different resolution levels, allowing a clear analysis of the resolution effects over the simulations. The TNG model includes new physical processes such as magneto-hydrodynamics, as well as new prescriptions for black hole formation, growth and multimode feedback, stellar evolution and chemical enrichment \citep{Pillepich18b} and is able to form smaller galaxies comparable with observations, with higher resolution level. In this work, we employ the same cosmology as employed by TNG on the parameters derived from \citet{Planck16} in which $\Omega_m=$ 0.31, $\Omega_\Lambda=$ 0.69, $\Omega_b=$  0.0486 and $h=$ 0.677.

\subsection{Sample construction}
\label{sec:sample_const}

\begin{table*}
    \centering
    \begin{tabular}{c c c c c c}
         & & Rotation-dominated & Dispersion-dominated & All morphologies \\ 
        \hline
        LSBGs & Centrals  & 2,454 &   777 &  3,231 \\
         & Satellites     &   891 & 1,685 &  2,576 \\
         & All            & 3,345 & 2,462 &  5,807 \\
        \hline
        HSBGs & Centrals & 4,838  & 4,151 &  8,989 \\
         & Satellites    & 2,999  & 4,744 &  7,743 \\
         & All           & 7,837  & 8,895 & 16,732 \\
    \end{tabular}
    \caption{Number of galaxies found in our galaxy sample, segregated by morphology (rotation-dominated or dispersion-dominated) and galaxy type (central or satellite).}
    \label{tab:galaxy_sample}
\end{table*}

In this work, we employed the simulated galaxy sample of \cite{PerezMontano22}, which makes use of the TNG100 run at its highest resolution level, consisting of (1820)$^3$ particles at the snapshot corresponding to $z=0$. This reproduces a galaxy sample consisting of objects comparable with ``local'' observations. The simulated galaxy sample includes galaxies with $10^{9} < $ log(M$_{*}$) $< 10 ^{12} \Msun$. For a given subhalo, we calculate the specific angular momentum of the stellar component $\vec{j_{*}}$ as

\begin{equation}
	\label{eq:angmom}
	\vec{j_{*}} = \frac{\vec{J}}{M} = \frac{\sum_i m_i \vec{r_i} \times \vec{v_i}}{\sum_i m_i},
\end{equation}

\noindent where $\vec{r_i}$ and $\vec{v_i}$ are the position and the velocity vectors of each stellar particle with mass $m_i$ within the given subhalo. The azimuthal component of $\vec{j_{*}}$ is employed to project the position of all the stellar particles over the axis which is parallel to it. We then calculate the projected effective radius $R_{50,r}$ which contains half of the total luminosity, by applying a linear interpolation over the cumulative distribution of the $r$-band luminosity, as a function of the projected 2D-distance $r$ of each stellar particle with respect to the center of the subhalo. Luminosities in TNG are obtained following a single stellar population model given by \citet{BC03}. This allows us to obtain the apparent magnitude $m_r$ (assuming that galaxies are located at $z=0.0485$) considering the total flux of the stellar particles enclosed by $R_{50,r}$, which is employed to finally obtain the central surface brightness $\mu_r$ within $R_{50,r}$ following \citet{Zhong08} and \citet{Bakos12} as

\begin{equation}
	\label{eq:mu_x}
	\mu_r = m_r +2.5 \log{(\pi R_{50,r}^2)}.
\end{equation}

We applied this methodology over all the galaxies in TNG100 with $10^{9} < $ log(M$_{*}$) $< 10 ^{12} \Msun$, and LSBGs are classified according to their central surface brightness if $\mu_r > 22.0$ mag arcsec$^{-2}$; otherwise, galaxies are classified as HSBGs. Our full sample consists of 22,539 galaxies, including 5,807 LSBGs and 16,732 HSBGs.

In order to distinguish between rotation-dominated and dispersion-dominated galaxies, we consider a kinematic criterion $\kappa_{rot}$ \citep{Sales10,RodGom17}, defined as the ratio between the rotational energy $K_{rot}$ from the azimuthal component of the stellar velocities, and the total kinetic energy $K$, i.e. 

\begin{equation}
\label{eq:kappa}
	\kappa_{rot} = \frac{K_{rot}}{K}.
\end{equation}

It is worth mentioning that, hereafter, throughout this work we will refer as ``rotation-dominated'' to those galaxies with the presence of a disk, regardless of the presence of arms and  spiral structure, given that we are interested in those galaxies with a component that can be modeled by an exponential profile, supported by rotation. Rotation-dominated galaxies in our sample correspond to galaxies with $\kappa_{rot} > 0.5$, while dispersion-dominated are those with $\kappa_{rot} < 0.5$. Table \ref{tab:galaxy_sample} presents a summary of the galaxy sample, segregated by morphology. Fig. \ref{fig:images} shows a set of synthetic images of four randomly selected galaxies in our sample with an stellar mass of $M_{*} \sim 10^{10.5} \Msun$ with different morphologies and surface brightness. The top (bottom) row illustrates a face-on (edge-on) view of the selected galaxies. First and second columns correspond to rotation-dominated galaxies, while the third and fourth columns are dispersion-dominated. These images were obtained by following the prescription of \citet{RodGom19} applied to the $g,r,i$ bands of SDSS.

\subsection{Supplementary data}
\label{sec:suplementary_data}

First, to characterize the environment of LSBGs at intermediate scales, we consider the overdensity field $\delta$ around the galaxies in our sample, which is based on the distance to the fifth nearest neighbour with an $r$-band magnitude brighter than -19.5, following the methodology of \citet{Vogelsberger14} and \citet{RodGom16}.

Second, to perform a full study of the environment at larger scales, we explore the spatial distribution of LSBGs with respect to the large-scale structure of the Universe. The large-scale distribution of the galaxies in our sample is drawn from \citet{Duckworth20a} and \citet{Duckworth20b}, which provide a post-processed identification of the topology of the cosmic web employing the Discrete Persistent Structure Extractor code (\texttt{DisPerSE}, \citealt{Sousbie11a,Sousbie11b}), in which the authors identify topological structures such as peaks, voids, walls and filaments within sampled distributions in 2D or 3D\footnote{Details in \url{http://www2.iap.fr/users/sousbie/web/html/indexd41d.html}}. \texttt{DisPerSE} uses a set of discrete points (or galaxies) to estimate the density field and the cosmic web. In this catalog, each point corresponds to a galaxy with $M_{*} > 10^{8.5} \Msun$, and the code is designed to be able to recover the galaxy distribution from an observational survey. This catalog provides the cosmic web distances to the nearest cosmic web structure of a given type:\\

	i) \textit{Void:} Minimum critical point.\\
	\indent ii) \textit{1-saddle point:} The critical point where one dimension is collapsing.\\
	\indent iii) \textit{2-saddle point:} The critical point where two dimensions are collapsing.	\\
	\indent iv) \textit{Node:} Maximum critical point.\\	
	\indent v) \textit{Filament:} Segment midpoint.\\
	
By definition, every filament starts and ends at a node and saddle points are minima along the filament. The reconstruction of the cosmic web is performed by using two different sets of tracers, the first, considering all subhalos that contain a minimum stellar subhalo mass of ${\displaystyle 10^{8.5} \Msun}$, and the second, using all subhalos that contain a minimum total subhalo mass (including DM and gas masses) of ${\displaystyle 10^{8.5} \Msun}$. For this work we employ the one performed considering only stellar particles. In the following sections, we performed a detailed analysis on the environment of LSBGs at different scales, starting from smaller scales comparable with group sizes, following the local overdensities, and finally reaching the large scale structure of the universe.

\begin{figure*} 
    \centering
    \includegraphics[width=0.75\textwidth]{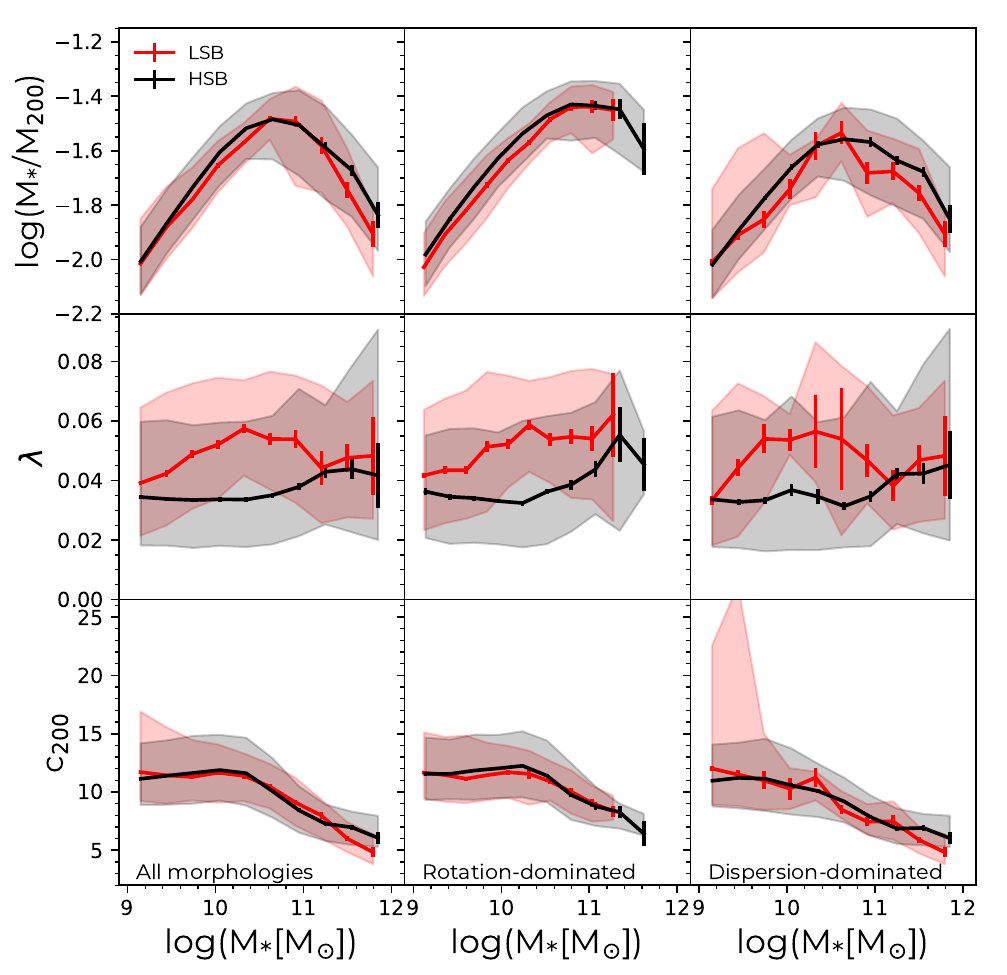}
    \caption{Median values of main halo properties hosting central LSBGs and HSBGs, which include the stellar-to halo mass fraction (top row), the spin parameter (middle row) and halo concentration index (bottom row). At fixed stellar mass, the main differences between both galaxy populations are found in the values of $\lambda$, regardless of their morphology. This suggests that the formation of LSBGs is more likely to be caused by variations in the angular momentum of their host halos, rather than variations in their density (characterized by $c_{200}$), as one of the mechanisms highlighted by \citet{McGaugh21}.}
    \label{fig:c200}
\end{figure*}

\section{Results}
\label{sec:Results}

The results found in \citet{PerezMontano22} suggest that the high angular momentum and stellar mass accretion phenomena are most likely to be the main mechanisms that guide the formation of an extended component and therefore, being responsible for the observed properties that LSBGs present in the current epoch. A detailed analysis of the environment around these galaxies at different scales would allow us to test such hypothesis, revealing clues on which of the mechanisms have the most impact on the evolution of the galaxies in our sample. The natural starting point from our study includes the small scales comparable with galaxy groups and their hosting DM halo (sec. \ref{sec:halo}). We then move up to group scales, allowing us to explore the overdensity field around these galaxies (secs. \ref{sec:groups} and \ref{sec:overdens}), to finally complete our study with the largest structures of the cosmic web (sec. \ref{sec:lss}), implementing different statistics that allow us to characterize the environment at different scales. Although these statistics most certainly present strong correlations, the variance among them is significant, and their implementation will provide a complete picture. In this section, our main goal is to determine the role that the environment plays in the formation and evolution of LSBGs in our simulated sample.

\subsection{Intrinsic Halo Properties}   
\label{sec:halo}

For this specific section, we will consider only `central' galaxies to perform our analysis, due to the fact that most of the properties studied here are well-defined only for central galaxies, which are usually  the most massive galaxy of their parent FoF group. 

Previously in \citet{PerezMontano22}, we found that LSBGs are formed within halos of marginally smaller stellar-to-halo mass ratio, but with higher spin parameter than HSBGs. We reproduce this result presenting the median values of $M_{*}/M_{200}$ and $\lambda$ as a function of $M_{*}$ in the first column of the upper and middle rows in Fig. \ref{fig:c200}. The red and black lines indicate the median values found for LSBGs and HSBGs, respectively, and the shaded area encloses the region delimited by the 16th and 84th percentiles of the corresponding distributions. Errorbars represent the statistical significance of the median, and are obtained by performing a Bootstrap re-sampling with 1000 random realizations derived from the original samples. Upcoming plots will follow the same format throughout this manuscript. In this figure, we extended our previous analysis by separating the galaxies by morphology, according to their $\kappa_{\rm rot}$ values. We visualize this segregation in the 2nd and 3rd columns of Fig. \ref{fig:c200}. We observe that, despite their morphology, we find similar trends when compared with the full sample.

These differences are not surprising in the case of rotation-dominated galaxies, due to the fact that many observational studies found that LSBGs are mainly disk-dominated systems, and therefore the study of their angular momentum and spin parameter indicates that LSBGs are galaxies with high angular momentum (\citealt{McGaugh95a,Zhong08,Galaz11,PerezMontano19,Salinas21}). The interesting fact is that, in our sample, even dispersion-dominated galaxies are found to have higher values of their spin parameter when compared to HSBGs. This trend might be due to the inclusion of galaxies with $\kappa_{rot}=0.4-0.5$, which roughly corresponds to lenticular galaxies rather than purely dispersion-dominated systems. This suggests the presence of a rotation component, even in galaxies with $\kappa_{rot}<0.5$, as discussed in sec. \ref{sec:sample_const}.

\subsubsection{Halo concentration} 
\label{sec:c200}
In a recent review by \citet{McGaugh21}, it is pointed out that the variations in the size of the galaxies, and consequently the formation of LSBGs, could be followed from one of two effects: variations in the spin parameter of the halo, or variations in the density of the DM halos. The three panels of the middle row of Fig. \ref{fig:c200} show a clear difference in the median values of the spin parameter, where LSBGs exhibit, across a wide range of $M_{*}$,  higher values of $\lambda$ when compared with HSBGs, at fixed stellar mass, which may suggest that the first mechanism highlighted by \citet{McGaugh21} is the one that induces the formation of LSBGs in our sample. At $M_{*} \gtrsim 10^{11.5}$, these differences are no longer statistically significant. However, it is not possible to affirm that this is ``the main'' mechanism without studying the second scenario. To do so, in the lower row of Fig. \ref{fig:c200} we plot the median values of the halo concentration index $c_{200}$ of our synthetic sample. This concentration index in our sample is computed as
    
    \begin{equation}
        c_{200} = \frac{R_{200}}{R_s},
    \end{equation}

\noindent where $R_{200}$ is the radius in which the enclosed average density is 200 times the critical density of the Universe, and  $R_s$ is computed by fitting a Navarro-Frenk-White (\citealt{NFW96}) profile to the DM density profile. The values of $c_{200}$ are available on the post-processed data catalogs from \citet{Anbajagane22}. We observe no significant differences in the median values of $c_{200}$ between LSBGs and HSBGs, even if we segregate our galaxies by morphology, therefore, we can conclude that for the galaxies in our sample, the differences between LSBGs and HSBGs are consequence of variations in the spin parameter of the hosting halos, and are not a result of differences in the concentration of the corresponding halos. 

This is in good agreement with \citet{PerezMontano22}, where we tracked back the evolution of the spin parameter of our simulated galaxies, finding a causal relation between the variations of $\lambda$ and the changes of other properties, such as the size of our galaxies. Interestingly, our findings related to $c_{200}$ seem to be in contrast with \citet{Kulier20}, where the authors found that LSBGs tend to inhabit slightly more concentrated halos than HSBGs. They argued that this trend is probably a consequence of the strong correlation found in the EAGLE sample between the concentration index and the mean stellar age of the central galaxy and its star formation rate (\citealt{Matthee17,Matthee19}). Given that in \citet{Kulier20} the central galaxies of halos with larger concentrations are older, it is expected to find LSBGs residing in halos with high concentration index. However, in \citet{PerezMontano22} we found no significant difference in the stellar ages of central galaxies between LSBGs and HSBGs, and consequently, no differences in the concentration indices of halos in TNG100 were expected.

\begin{figure} 
    \centering
    \includegraphics[width=0.4\textwidth]{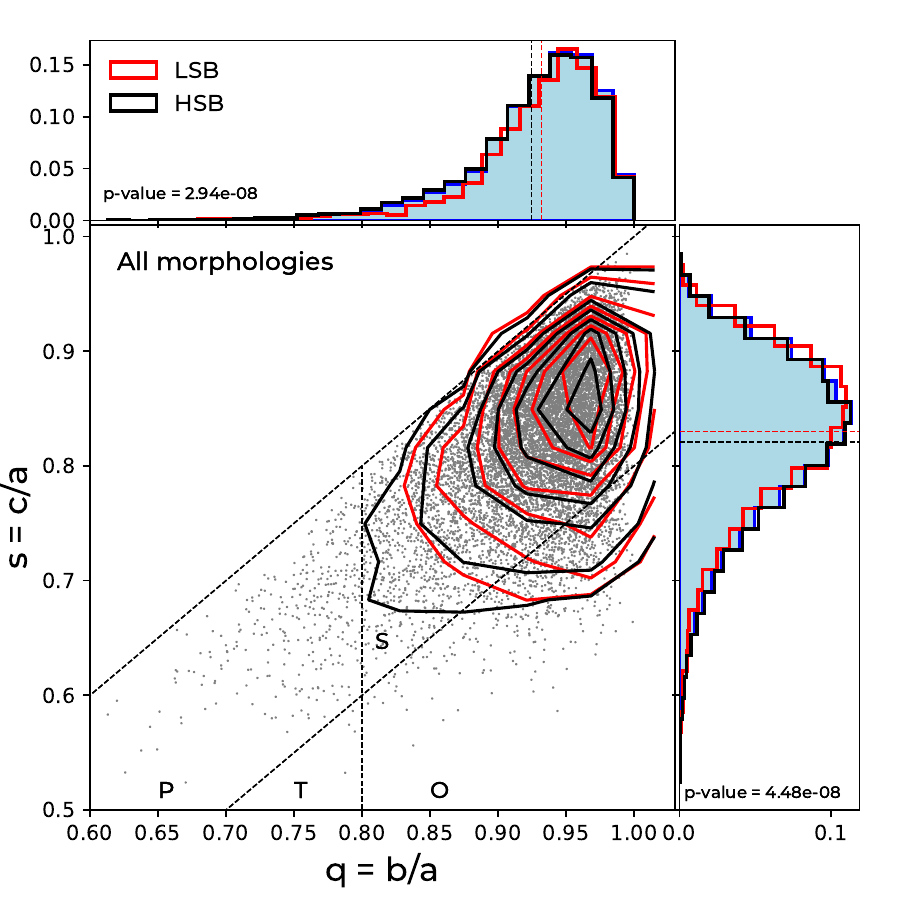}\\ \includegraphics[width=0.4\textwidth]{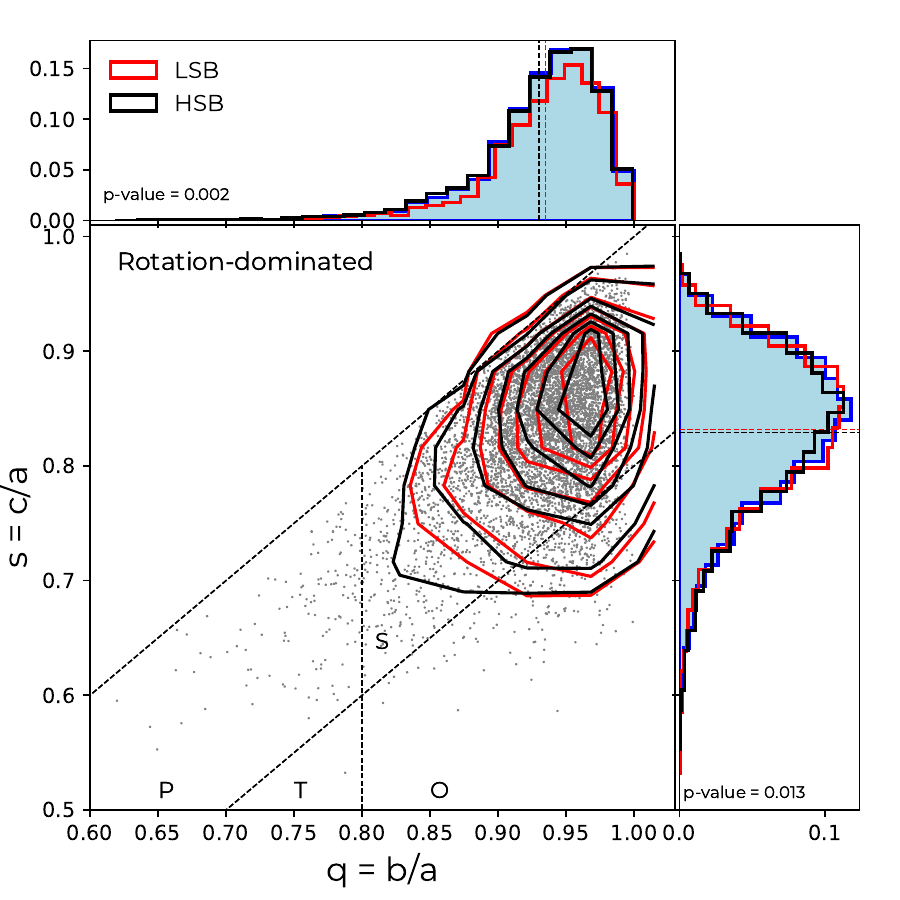}\\ \includegraphics[width=0.4\textwidth]{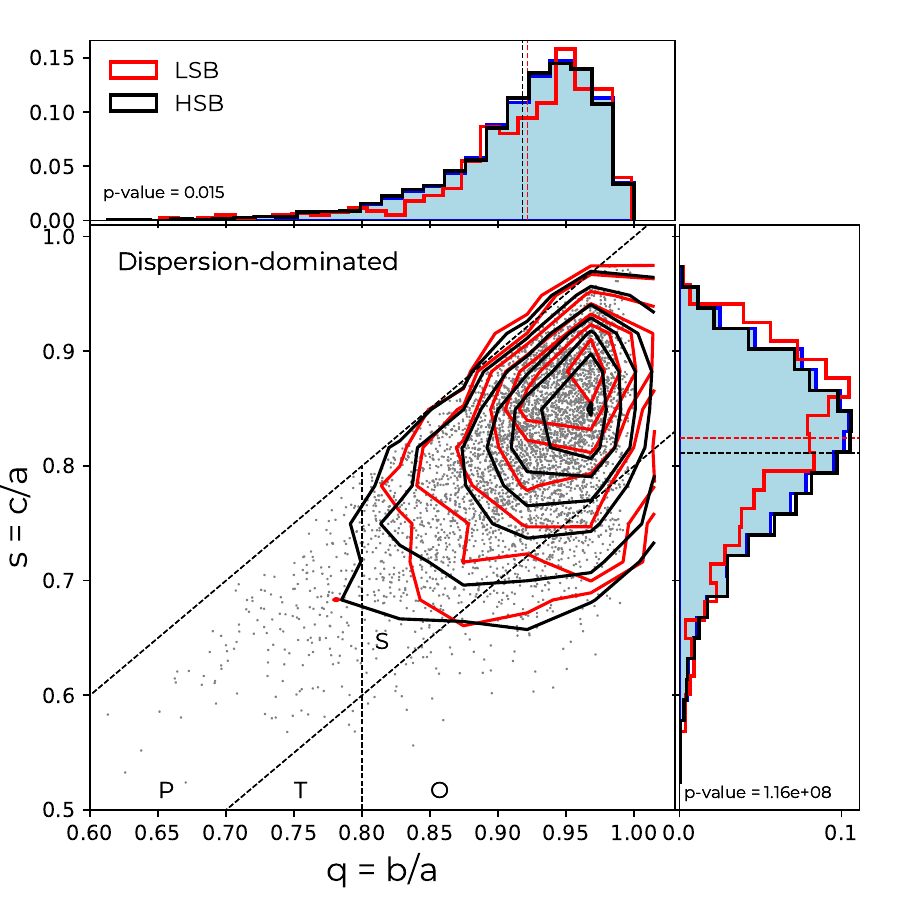}
    \caption{The axis ratios of the DM halos for all galaxies in the sample (top), rotation-dominated (middle) and dispersion-dominated (bottom). Dotted lines indicate the regions where the halo shapes are considered as prolate (P), tri-axial (T), oblate (O) and spherical (S). We can observe that not only LSBGs, but also HSBGs, are found in nearly spherical halos, and the departure from spherical halos in HSBGs is larger than in the case of LSBGs.}
    \label{fig:axis_full}
\end{figure}

\subsubsection{Halo Shape}
\label{sec:shapes}

The study of the halo shape can provide interesting clues regarding different processes involved in the formation and evolution of galaxies. We continue the study of the dark matter halo configuration by analyzing the shapes of the halos in which LSBGs and HSBGs reside. Fig. \ref{fig:axis_full} shows the intermediate-to-major axis ratio $q=b/a$ and the minor-to-major axis ratio  $s=c/a$ of the dark matter halos, which are taken from the supplementary catalog of \citet{Anbajagane22}. Dotted lines delimit the regions in which the halos are classified as \textit{triaxial} ($a > b > c$), \textit{prolate} ($a > b = c$), \textit{oblate} ($ a= b > c$) and \textit{spherical} ($ a = b = c $). Density contours show that most of the galaxies in our sample (around 91\% of them) present values of $q$ and $s$ closer to 1, implying that most of our galaxies reside within nearly-spherical halos, regardless of their morphology. By contrast, only 6\% of the galaxies reside within halos with oblate geometry, while 3\% live within prolate halos, and only 4 galaxies in our sample are found in triaxial halos. We also plot in Fig. \ref{fig:axis_full} the distributions of $q$ and $s$ along the corresponding axes, segregated by LSBGs and HSBGs. We found a small but statistically significant difference in the distributions of $q$ and $s$, where both axis ratios in LSBGs are closer to 1 than in the case of HSBGs. The figure also includes the corresponding $p$-values obtained from a Kolmogorov-Smirnov (KS) test performed over LSBGs and HSBGs distributions. In all the cases, the $p$-values $<< 0.01$ allow us to reject the null hypothesis that these were extracted from the same population. 

To ensure that the observed trends are not influenced by the mass of the galaxies, we conducted a comparison with three distinct sets of control samples. These control samples were created by selecting a LSBG from our original sample and identifying a HSBG counterpart with a difference in stellar mass of $\Delta$ $\log M_{*}$ $< 0.05$. This criterion was employed to ensure statistical similar mass distributions between both sets. We established three pairs of control samples: one encompassing all central galaxies in our sample, irrespective of their morphology (`CS1'), and two additional pairs segregated into rotation-dominated (`CS2') and dispersion-dominated galaxies (`CS3'). The selection of these pairs ensured an equal number of galaxies in each control sample. Subsequently, a KS-test was conducted for each couple of control samples to validate that their stellar mass distributions are statistically indistinguishable. 

\begin{figure*} 
    \centering
    \includegraphics[width=0.95\textwidth]{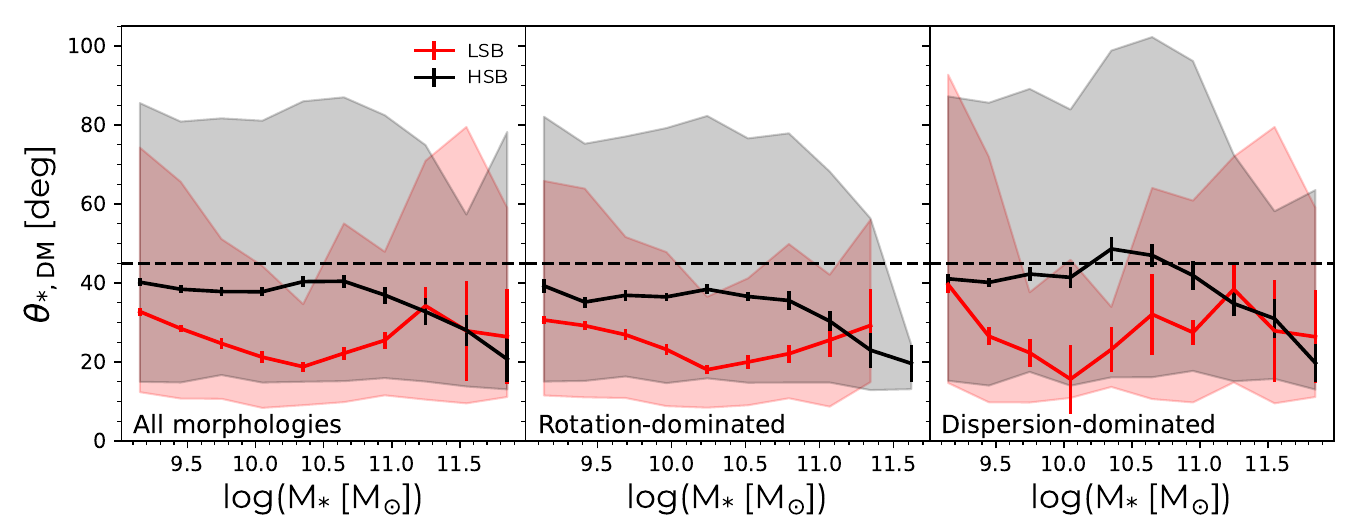} 
    \caption{The alignment between stellar and halo angular momentum vectors, represented by the angle between them. The distributions of $\theta_{*,\rm DM}$ are found to be narrower for LSBGs than the ones obtained for HSBGs, with median values closer to zero, indicating a stronger alignment between components in LSBGs than in HSBGs.}
    \label{fig:theta}
\end{figure*}

The results obtained for the geometry of the halos of the galaxies in each pair of control samples are very similar to the ones obtained for the full sample (without controlling the stellar mass), that is, most of the galaxies reside within quasi-spherical DM halos, with LSBGs showing $q$ and $s$ values closer to 1 than HSBGs.

\subsubsection{Angular Momentum Alignment}
\label{sec:alignment}

Previous studies employing numerical simulations have shown the existence of an alignment between stellar and dark matter angular momentum vectors (e.g. \citealt{Bailin05,Bett10,Velliscig15}). The existence of such an alignment is the result of a natural combination of two phenomena: the common origin and co-evolution of a protogalaxy and its corresponding hosting halo, and the dynamical response of the halo to the assembly of the baryonic disk. Therefore, it is important to explore in detail the alignment of the angular momentum vectors between the stellar and dark matter components,providing some interesting clues towards the understanding of the galaxy/halo evolution, as well as the processes involved in it. 

With this premise, we compute the angular momentum of the dark matter component $\vec{j}_{200}$ employing a slightly different version of eq. \ref{eq:angmom}, this is
    
\begin{equation}
\label{eq:j200_def}
	\vec{j}_{200} = \frac{\vec{J}_{200}}{M_{200}} = \frac{\sum_i m_{DM} \vec{r_i} \times \vec{v_i}}{\sum_i m_{DM}}.
\end{equation}   

This equation provides the main halo's angular momentum $\vec{j}_{200}$ by considering, not the stars, but the DM particles within $R_{200}$. Similarly to the case of the stellar component, each dark matter particle of the configuration has associated to it a position and a velocity vector, and the mass of each dark matter particle is the same for all the particles considered in this calculation, with a value of $7.5 \times 10^{6} \rm{M}_{\odot}$. Equations \ref{eq:angmom} and \ref{eq:j200_def} provide the specific angular momentum vector for the stellar and dark components of the galaxies, from which we compute the alignment angle through 
   
\begin{equation}
        \label{eq:theta}
        \cos (\theta_{*,\rm DM}) = \frac{\vec{j}_{*} \cdot \vec{j}_{200}}{\|\vec{j}_{*}\| \|\vec{j}_{200}\|},
 \end{equation}

\noindent where $\theta_{*,\rm DM}$ is the angle between both angular momentum vectors $\vec{j}_{*}$ and $\vec{j}_{200}$. The left-hand panel of Fig. \ref{fig:theta} shows the median values of $\theta_{*,\rm DM}$ as a function of the stellar mass, considering all the central galaxies in our sample, and the dashed line highlights the 45$^{\circ}$ angle between the angular momentum vectors of both components. From this, it can observed that the median values of $\theta_{*,\rm DM}$ are systematically lower for LSBGs than in the case of HSBGs, implying a stronger alignment between both components than for HSBGs. 

This trend is qualitatively similar even if we divide our sample into rotation-dominated (middle panel) and dispersion-dominated (right-hand panel) galaxies. Interestingly, in the three panels of Fig. \ref{fig:theta} we observe that the median values of $\theta_{*,\rm DM}$ are found mostly under the 45$^{\circ}$ line, with the corresponding $\theta_{*,\rm DM}$ distributions in LSBGs being narrower than for the case of HSBGs. This result highlights the role of the angular momentum alignment in galaxy evolution, such that aligned components favor a scenario in which the total angular momentum is highly conserved, causing both vectors to be added in the same direction, leading to the formation of LSBGs, while a misalignment would imply that most of the total specific angular momentum will be cancelled due to the angular momentum components pointing in different directions. However, we must point that additional factors may play even a more central role, such as the intrinsically higher spin in LSBGs (Fig. \ref{fig:c200}, middle row), differences in the angular momentum retention fraction \citep{PerezMontano22,RodGom22} and additional complex hydrodynamical processes \citep{Genel15,Trapp24}. 

The angular momentum alignment between different components has been found to be important in the formation mechanisms of LSBGs, as highlighted by \citet{DiCintio19}, where the authors found that, aside from mergers, the alignment of the angular momentum of accreting gas through cosmic time is a key factor on the determination of the final galactic morphology, such that LSBGs are formed out of gas flowing in with similar angular momentum as the previously accreted material \citep{Zhu23}. Related to the alignment of the stellar and dark components, \citet{Bullock01} showed that halos with a misaligned angular momentum distribution may be unlikely to host extended galaxies, as are the LSBGs in our sample \citep{PerezMontano22}.

\subsection{Local Effects within Groups and Clusters. }
\label{sec:groups}

\begin{figure*} 
    \centering
    \includegraphics[width=0.95\textwidth]{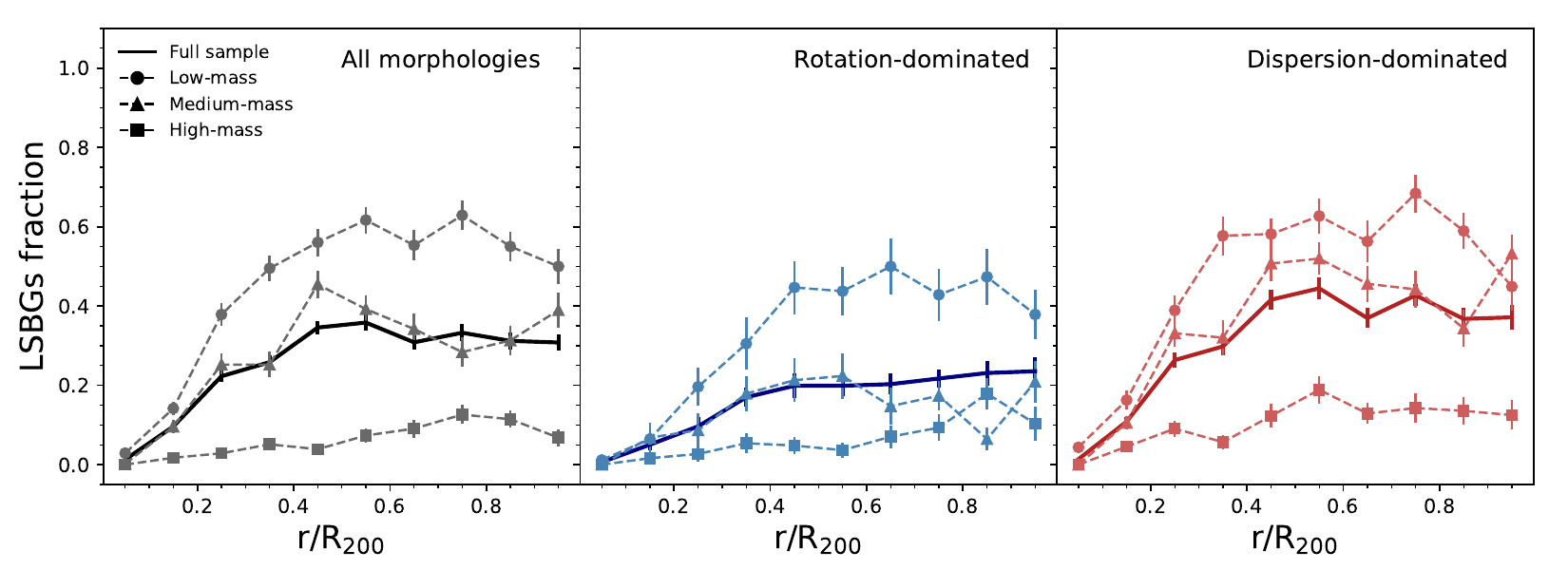} 
    \caption{The fraction of LSBGs as function of $r/R_{200}$, segregated by morphology within three different mass intervals. We observe that the fraction of LSBGs increases as we move away from the center of galaxy groups. At $r/R_{200} > 0.4$, the fraction of rotation-dominated LSBGs is nearly half of the fraction of dispersion-dominated ones, suggesting that close interactions may have transformed them into dispersion-dominated systems, implying that rotation-dominated and dispersion-dominated LSBGs may have followed different formation mechanisms.}
    \label{fig:groups}
\end{figure*}

We now turn our attention to the analysis of the environment around LSBGs at the scale of groups and clusters. As mentioned before, this is important because several authors have studied the environment of LSBGs at this level, in order to quantify how isolated these galaxies are \citep{Bothun93,Rosenbaum09,Galaz11,PerezMontano19}. 

According to simple theoretical considerations, the density of galaxies in a given group/cluster is expected to increase towards the center. However, such arguments typically neglect dynamical processes within group or cluster environments, such as tidal stripping and galaxy harassment, which are modeled self-consistently in cosmological simulations. With these considerations, in order to study the impact of the groups in which LSBGs reside on their observed properties, we quantify the fraction of satellite LSBGs as a function of the distance $r$ to the center of its corresponding host group. Consistent with the \texttt{SUBFIND} convention, the position of each galaxy (central or satellite) corresponds to the coordinates of its most gravitationally bound particle. Once identified the central galaxy and the rest of the satellite galaxies, we calculate the 3D distance from the center of the corresponding satellites to the central one. 

Due to the wide distribution of halo sizes in our sample, we will focus on the ``relative'' distance to the center of a given group to obtain a better understanding of our results. To do so, we normalize the computed distance by the characteristic ``group size'' $R_{200}$, defined as the radius containing 200 times the critical density. We perform this experiment for every group in our sample (i.e., with at least two galaxies), and finally obtain the net fraction of LSBGs within different $r/R_{200}$ bins.

In Fig. \ref{fig:groups} we plot the net LSBGs fractions as function of the normalized clustercentric distance $r/R_{200}$. This figure includes satellite galaxies of all morphologies (left-hand panel), as well as segregated into rotation-dominated (middle panel) and dispersion-dominated (right-hand panel). We segregate our sample by stellar mass by ranking all the satellite galaxies by $M_{*}$, and then dividing the full distribution in three sub-samples with the same number of galaxies, each of them having different median values of $M_{*}$ (named `low', `medium', and `high'). This procedure was chosen, rather than imposing lower and upper limits of stellar mass in every sub-sample, because the number of LSBGs at stellar masses larger than $M_{*} = 10^{11.5} \rm M_{\odot}$ in our complete simulated sample is quite small, biasing the interpretation of our results at those stellar mass ranges. Different line-styles indicate these different mass ranges, while solid lines include all the galaxies considered in the corresponding analysis. 

From the solid line in the left-hand panel of Fig. \ref{fig:groups}, when no morphological segregation is introduced, we note that most of the satellite LSBG population is further away from the center of galaxy clusters, being preferentially  found in the outer regions of such clusters. This might indicate that the mechanisms that might transform HSBGs into LSBGs (such as tidal stripping and stellar accretion) are not efficient in the central regions, or they are destroyed once they reach the central regions, confirming that LSBGs tend to reside in low-density environments. Focusing on the segregation by stellar mass (dotted-lines), we note that the fraction of LSBGs is larger for low-mass galaxies (circles) than in the case of high-mass galaxies (squares). For low-to-intermediate mass galaxies, we clearly see that the fraction of LSBGs depends on the distance to the center of the galaxy cluster, as there is an absence of LSBGs close to the center. In the case of high-mass galaxies, such dependency is weaker, suggesting that these galaxies are massive enough to ``survive'' violent events. 

However, when we segregate our sample between rotation-dominated and dispersion-dominated galaxies, the fraction of rotation-dominated LSBGs decreases faster than in the case of dispersion-dominated, implying that LSBGs at the center of galaxy clusters are more likely to be of early type. The fraction of rotation-dominated LSBGs is nearly constant at 20\% for $r/R_{200} > 0.4$, while for dispersion-dominated LSBGs this fraction is almost double. Moreover, the fraction of LSBGs of intermediate masses is larger for dispersion-dominated galaxies than rotation-dominated ones. 

The observed outcome not only suggests a strong dependency of LSBGs formation on their proximity to the center of a given group/cluster and on their morphology, but also proposes a plausible explanation for the absence of intermediate-mass rotation-dominated LSBGs. This absence may signify the susceptibility of these galaxies to interactions, potentially leading to an overabundance of satellite dispersion-dominated LSBGs within our sample, as highlighted by \citet{Impey97} and \citet{DiCintio19}, who underscore the likelihood of diverse formation scenarios for LSBGs: rotation-dominated LSBGs might have originated from secular evolution or minimal exposure to disruptive interactions, while dispersion-dominated LSBGs are more likely to result from varied mechanisms induced by intense galaxy interactions, such as mergers (similar to the findings in \citealt{Kulier20}).

\subsection{Large-Scale Environmental Effects}
\subsubsection{Overdensity}
\label{sec:overdens}

\begin{figure*}
    \centering
    \includegraphics[width=0.85\textwidth]{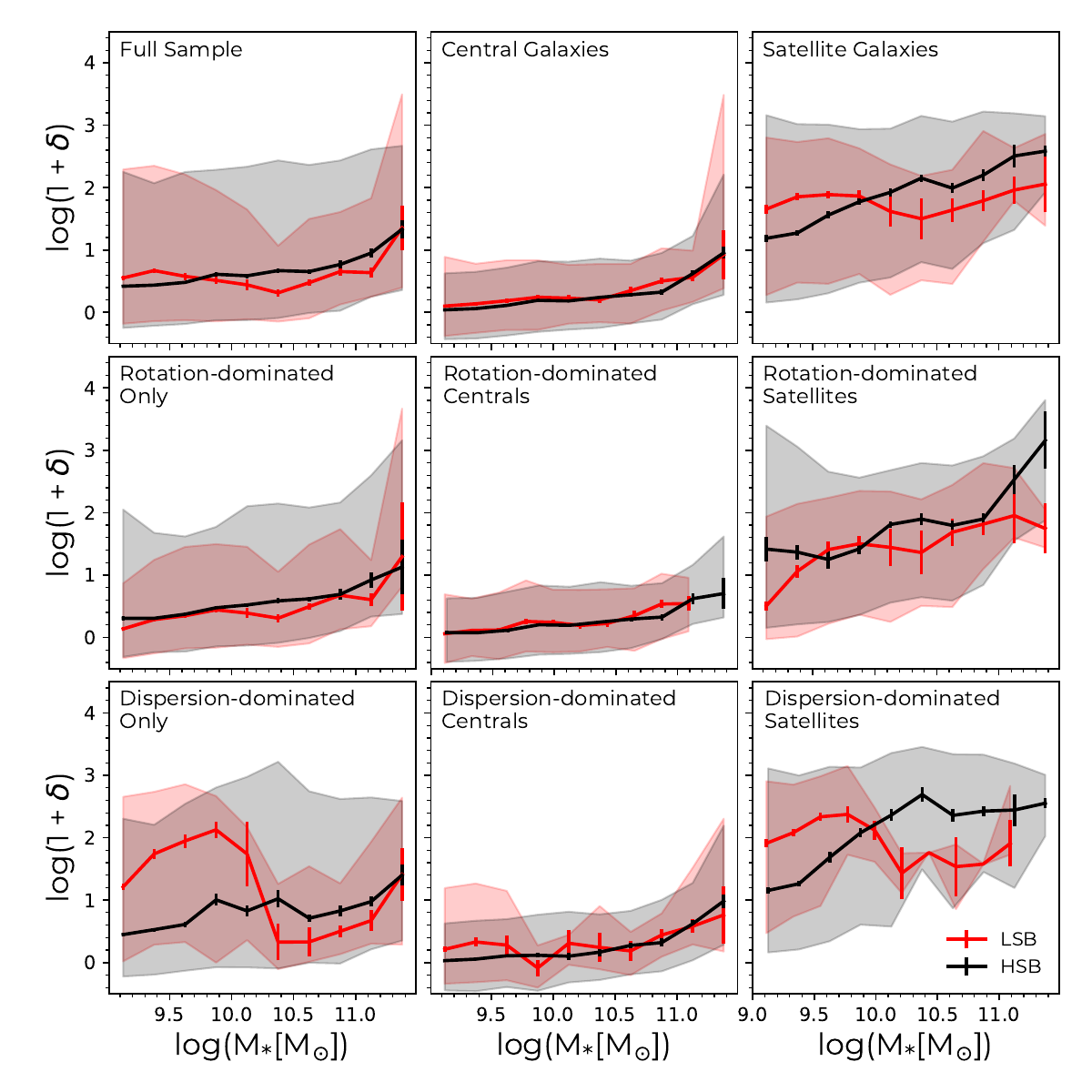} \\
     \caption{Median values of the overdensity field $\delta$ as function of $M_{*}$ of our simulated sample. Each panel corresponds to a different subset of galaxies segregated by galaxy-type (central or satellites) and by morphology. We found no significant difference in the overdensity field between central LSBGs with rotation-dominated morphology, which could be considered as an indication of secular evolution. Given that galaxies in high-density environments are most likely to have many interactions, this could indicate that satellite dispersion-dominated LSBGs are formed mainly due to mergers and flybys.}
    \label{fig:density_contours}
\end{figure*}

 To fully understand the influence of galaxy companions, we present in Fig. \ref{fig:density_contours} the median values of the overdensity field $\log(1+\delta)$ as function of $M_{*}$. Each panel corresponds to a different subset of galaxies identified in our sample, that is, by galaxy type (central or satellite) and by morphology (rotation or dispersion dominated). The overdensity field was estimated by using galaxies with $r$-band absolute magnitudes brighter than $-19.5$ as tracers, and using the distance to the 5th nearest neighbour as a smoothing length \citep{Vogelsberger14, RodGom16}. From this figure, different environmental effects can be addressed for different sub-populations of galaxies.
 
For satellite galaxies with no morphology segregation (upper-right panel) we observe a transition in which, for low-mass galaxies, LSBGs are found in high density environments, where the distribution is dominated by the dispersion-dominated population (lower-right panel). However, the opposite behavior is observed in the high-mass regime, which is dominated by the rotation-dominated population. This difference suggests that the formation of rotation-dominated satellite LSBGs is likely favored in low-density environments, allowing them to preserve their low surface brightness nature. In contrast, dispersion-dominated satellite galaxies are found in regions of intermediate$/$high densities, indicating that the environment may play a more critical role in the formation of dispersion-dominated satellite LSBGs. 

It is important to note that, although LSBGs are found with different morphologies across various stellar mass ranges, massive satellite galaxies with dispersion-dominated morphology constitute an insignificant part of the LSBG population, as reported by \citet{PerezMontano22}. Moreover, for rotation-dominated systems, the number of central galaxies is larger than the number satellite ones, while for dispersion-dominated systems, the opposite occurs (see Table \ref{tab:galaxy_sample}), indicating that dispersion-dominated LSBGs are mostly linked to satellite populations. Therefore, the trends observed at $M_{*}>10^{10} \Msun$ for dispersion-dominated satellite galaxies should be interpreted with caution, as these results may not be representative due to the limited sample size. 

This result seems to be in line with previous studies focused on the main mechanisms that favor the formation of the so-called \textit{Ultra-Diffuse Galaxies} (UDGs) in low/high density environments (\citealt{Amorisco16,Sales20,Ogiya18,Benavides21,Marleau21,Jones21}) since similar mechanisms seem to occur in our sample of LSBGs. Moreover, the transition from disks to spheroids caused by interactions with the cluster via ram pressure stripping and tidal interactions \citep{Koda15}, could explain the lower fraction of rotation-dominated LSBGs found in groups observed in Fig. \ref{fig:groups}. \citet{Sales20} pointed out that the tidal stripping of the stellar components could drastically reduce their stellar mass, with their sizes remaining constant, such that galaxies that experience strong tidal stripping could be transformed into galaxies with lower surface brightness.

On the other hand, from Fig. \ref{fig:density_contours} we observe that for central galaxies (top-middle panel), there is no discernible trend in the values of $\log(1+\delta)$ between LSBGs and HSBGs, whether considering the entire sample or segregating it by rotation-dominated (central panel) or dispersion-dominated (bottom-middle panel) systems.

Taken together, our results indicate that rotation-dominated LSBGs are less exposed to interactions than dispersion-dominated systems, implying that the evolution of these latter is mainly the result of the high-density environments. These results are in good agreement with \citet{Zhu23}, in which they employed TNG100 to explore the origin of the extended disks found in `giant' LSBGs, finding  evidence that giant LSBGs are extended disks without any major mergers since $z=1$, indicating that these galaxies have not been disturbed for a long time on a cosmological scale. A similar result was found by \citet{Saburova22} with the aid of a simulated sample of `giant' LSBGs drawn from EAGLE. Therefore, the relatively isolated environment is, therefore, one of the key parameters that explain the formation mechanism and `survival' of extended disks.

\begin{figure*} 
    \centering
    \includegraphics[width=0.85\textwidth]{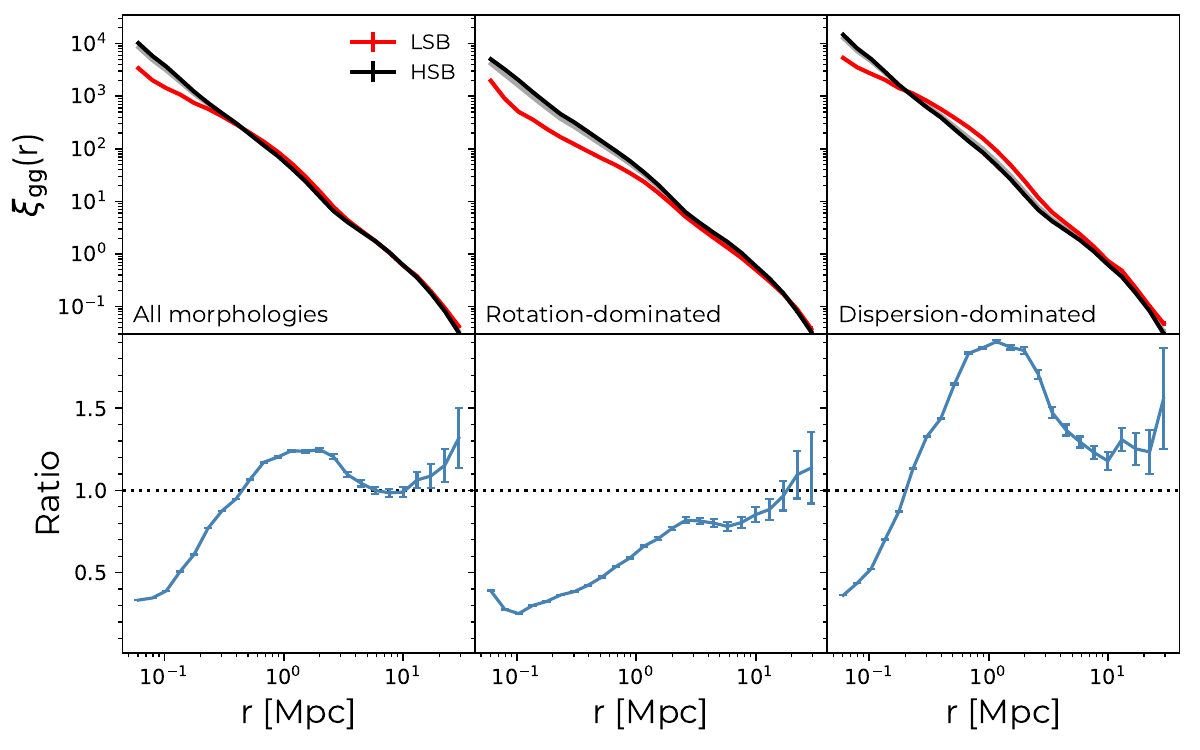} 
    \caption{\textit{Top row:} The two-point correlation function of LSBGs (red) HSBGs (black) and both (gray) with respect to the Full-TNG100 galaxy population. Error bars represent the dispersion around the values obtained for $\xi(r)$, computed by employing a boostrap re-sampling method. \textit{Bottom row}: The LSBGs-to-HSBGs ratio of the corresponding values of $\xi(r)$ on each bin. When no morphological segregation is considered (left-hand panels), LSBGs seem to be less clustered than HSBGs up to $r \lesssim 0.4$ Mpc, where the trend is inverted. However, when looking at galaxies of similar morphology, rotation-dominated LSBGs are found to be systematically less clustered than HSBGs at all scales, while the opposite is found for dispersion-dominated ones. The relative isolation of rotation-dominated LSBGs seems to be a key factor in the `survival' of their low surface brightness nature. Conversely, dispersion-dominated LSBGs are most likely to be a product of the interaction with other galaxies.}
    \label{fig:tpcf_full}
\end{figure*}

\subsubsection{Two-point correlation function}
\label{sec:tpcf}

We characterize the spatial distribution of LSBGs in our sample by computing the `two-point correlation function' (tpcf) $\xi(r)$, which is an important tool to study the galaxy clustering by measuring the probability of finding a galaxy within a distance $r$ from a reference galaxy target \citep{Peebles80}. To do so, we employed the \texttt{Halotools} package \citep{Hearin17}\footnote{Available for download at \url{https://halotools.readthedocs.io/en/latest/index.html}}, an open-source specialized Python package that allows the building and testing of galaxy-halo connection models. The package is able of creating mock observations of a synthetic galaxy population, starting from dark matter halo catalogs drawn from a given cosmological simulation, allowing a direct comparison with real observations. We use the \texttt{tpcf} command from the \texttt{mock-observables} sub-package to calculate the real space two-point correlation function, with a \citet{DavisPeebles83} statistical estimator:

\begin{equation}
\label{eq:tpcf}
    1+\xi (r) = \frac{DD(r)}{DR(r)},
\end{equation}

\noindent where $DD(r)$ is the number of original sample pairs with separations equal to $r$ and $DR(r)$ is the cross-count sum of all pairs within an interval $\Delta r$, in which one pair member belongs to the original sample and the other is taken from a random data set. We apply the \texttt{tpcf} command over the spatial coordinates of the galaxies in our sample to calculate the ``cross-correlation'' function of both, LSBGs and HSBGs, with respect to the full sample of galaxies in TNG100, that is, all the simulated galaxy population found in TNG100 with a minimum stellar mass of $10^{8.5} \Msun$, regardless if they are included or not in the LSBGs/HSBGs samples employed in this work. Hereafter, this later will be refereed as ``Full-TNG100''.

The upper panels of Fig. \ref{fig:tpcf_full} show the resulting tpcf of LSBGs, HSBGs, and both (indicated as red, black and gray lines, respectively), with respect to the Full-TNG100 sample. The upper-left panel corresponds to the cross-correlation functions of all the galaxies in our sample, regardless of their morphology. For clarity, the LSBG-to-HSBG tpcf amplitude ratio is shown in the lower panels of Fig. \ref{fig:tpcf_full}. A clear transition can be observed at $r \lesssim 0.4$ Mpc, where LSBGs are less clustered than HSBGs, and beyond this value the trend is inverted. This result indicates that the former are found in relative isolation when compared with HSBGs at small scales, but at larger scales, LSBGs seems to be slightly more clustered. 

The spatial distribution of LSBGs has been previously studied by \citet{Bothun93} and \citet{Mo94}, who showed that LSBGs are less strongly clustered than HSBGs, and that both galaxy populations follow the same large scale structure, but HSBGs adhere more strongly to it. However, the difference found in the amplitude of the tpcf between both galaxy populations results from the inclusion of galaxies of all morphological types and the different formation channels between rotation-dominated and dispersion-dominated galaxies. Some previous studies (e.g., \citealt{Mo92,Zehavi11}) have made emphasis on the differences found in the correlation function between early-type and late-type galaxies, finding that early-type galaxies (dispersion-dominated galaxies) are in general more clustered than late-type galaxies (rotation-dominated galaxies).

The results shown in sec. \ref{sec:overdens} may suggest that the density of the local environment of LSBGs is statistically different for late-type and early-type galaxies, showing that the later are preferentially found in medium-density environments, while the former are most likely to be located in low-density environments. However, this conclusion should be taken carefully, given that this trend may not be true at larger scales, as suggested by the $r \lesssim 0.4$ Mpc end of the upper-left panel of Fig. \ref{fig:tpcf_full}. In order to perform a deeper analysis in our results, we plot in Fig. \ref{fig:tpcf_full} the tpcf of the galaxies in our sample, segregated between rotation-dominated (middle panel) and dispersion-dominated (right-hand panel) galaxies, showing clear differences between them. 

In the case of rotation-dominated galaxies, we can observe that LSBGs are systematically less clustered than HSBGs at all scales, which is an indication that these systems are, in general, more isolated than their HSBG counterparts. On the other hand, dispersion-dominated LSBGs exhibit a higher amplitude of the correlation function than HSBGs, implying that dispersion-dominated LSBGs are preferentially found within high-density environments, making them more susceptible to strong interactions that could play an important role in their formation. This provides key information in the interpretation of our results, which will be described in detail in the following section.

Given that the correlation function depends strongly on the stellar mass of the galaxies considered, we tested our implementation of the tpcf calculation over a new set of control samples obtained with the same procedure described in sec. \ref{sec:shapes}, but this time controlling not only by stellar mass, but also including both, central and satellite galaxies, in order to suppress any mass dependency in the interpretation of our results. We built three different pairs of control samples for this purpose: one for all the galaxies in our sample regardless of their morphology (`CS4'), and two more pairs of control samples segregated into rotation-dominated (`CS5') and dispersion-dominated galaxies (`CS6'). We performed a similar KS-test over each pair of control samples to test the hypothesis that the galaxies are drawn from the same distribution. 

When analyzing the tpcf within the control samples segregated by morphology, we observed strikingly similar outcomes (both qualitatively and quantitatively) compared to those depicted in Fig. \ref{fig:tpcf_full} for the entire sample. Consequently, we infer that the disparities evident in the cross-correlation function values of our galaxy sample primarily stem from the low surface brightness nature of these systems, rather than differences in their stellar mass.

\subsubsection{Distribution along the cosmic-web structure}
\label{sec:lss}

\begin{figure} 
    \centering
    \includegraphics[width=0.46\textwidth]{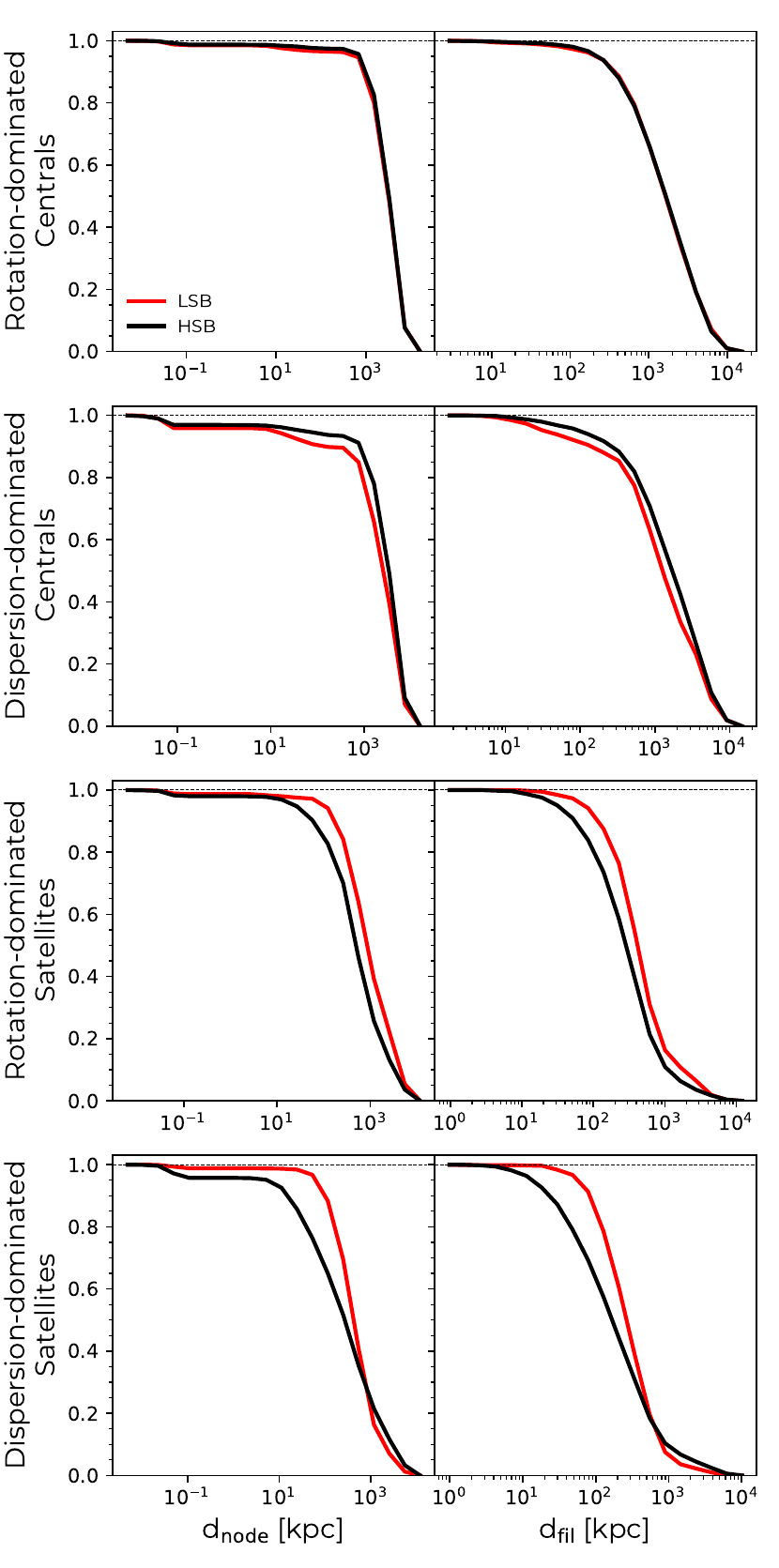}
    \caption{Cumulative distributions of the distance to the nearest large-scale structure for LSBGs and HSBGs, segregating the sample by central/satellite galaxies, and by morphology. The most noticeable differences along the cosmic web distribution of LSBGs are found in the distance to the closest node and filament, for central dispersion-dominated galaxies (second row), and satellite galaxies, regardless their morphology (third and fourth row).}%
    \label{fig:cumulative}
\end{figure}

We finally explored the distribution of our sample of LSBGs within the cosmic web. The cosmic web topology and galaxy distribution are based on \citet{Duckworth20a} and \citet{Duckworth20b}, employing the \texttt{DisPerSE} code for cosmic web topology characterization (\citealt{Sousbie11a,Sousbie11b}). This code identifies topological structures by employing a set of discrete points to estimate the density field and the cosmic web, and is designed to be able to recover the galaxy distribution from an observational survey. We employed this catalog to obtain the distances to the nearest cosmic web structure of a given type, namely voids, walls, filaments and saddle points\footnote{Further details on the identification of the large-scale structures and the theory behind the algorithm employed in \texttt{DisPerSE} is available on the corresponding papers of \citet{Sousbie11a} and \citet{Sousbie11b}}.

Fig. \ref{fig:cumulative} presents the cumulative distribution function of the distance to the closest of these large-scale structures for the galaxies in our sample, segregated between central/satellite galaxies, and their rotation-dominated/dispersion-dominated morphologies. This segregation is specified on the extreme left of the corresponding row. In most cases, we  found no significant difference in the spatial distribution of LSBGs and HSBGs, therefore, only included in Fig. \ref{fig:cumulative} the cases showing the most noticeable differences, namely the distance to the nearest node (left column) and nearest filament (right column). 
 
For central dispersion-dominated galaxies, we found that LSBGs are slightly closer to nodes than HSBGs (left column, second row). This result indicates that dispersion-dominated LSBGs are more likely to be found in high-density environments, where they are subject to many interactions, favoring a scenario in which dispersion-dominated LSBGs may have formed due to closer encounters with other galaxies, as well as the accretion of low-mass companions. In contrast, for central rotation-dominated galaxies we did not find a significant difference between the large-scale distribution of LSBGs and HSBGs (first row). This last result is in agreement with observational studies that found no strong impact of the cosmic web topology on the evolution of LSBGs, specially those with late-type morphology (e.g., \citealt{Mo94,PerezMontano19}). Interestingly, our findings allow us to extend such analysis to the full population of LSBGs, regarding their morphology.

In the case of satellite galaxies, rotation-dominated LSBGs are found to be further away from nodes than HSBGs (left column, third row) and, consequently, further away from the center of galaxy clusters. Similarly, rotation-dominated LSBGs are slightly further from nodes than dispersion-dominated LSBGs (left column, fourth row), which is consistent with what we observe in Figs. \ref{fig:groups} and \ref{fig:density_contours}, where most of these galaxies are preferentially located in low-density environments. On the other hand, dispersion-dominated satellite LSBGs are also found further from nodes than dispersion-dominated satellite HSBGs, but the former seem to be closer to nodes than rotation-dominated satellite LSBGs. This corroborates what we showed in Fig. \ref{fig:groups}, where the fraction of LSBGs increases as we move away from the center of galaxy clusters, and the fraction of dispersion-dominated LSBGs is larger at all radii. This is also in good agreement with the results in Fig. \ref{fig:density_contours}, where we found that the fraction of satellite dispersion-dominated LSBGs increases with density.

When looking at the distance to the nearest filament (right column), the most significant differences are found for satellite galaxies. We observe that LSBGs are located further away from filaments than HSBGs, regardless of their morphology. This is similar to what it was found by previous observational studies (\citealt{Rosenbaum09,PerezMontano19}). 

It has been shown that low-mass systems are preferentially located close to low-density environments such as voids \citep{Rojas05,Florez21}, implying that such dependency between stellar mass and density could strongly affect the interpretation of our results. In order to vanish any bias due to stellar mass, we performed a similar analysis over four new sets of control samples following the same procedure described in secs. \ref{sec:shapes} and \ref{sec:tpcf}. This time, apart from controlling by stellar mass, we also control by galaxy type, segregating between central and satellite galaxies, as well as morphology. The results of such experiment over the control samples are not significantly different from the ones obtained for the full sample. Therefore, we can conclude that the small differences found can be attached to the LSBG nature of these systems, rather than their distribution along the cosmic web.

The weak differences found in the large-scale distribution along the cosmic web between LSBGs and HSBGs is in good agreement with some recent studies \citep{Kuutma20,Das22} that highlight that the most immediate environment clearly has a greater impact on galaxy properties (such as color and star formation rates), than the large-scale structure in which the galaxies reside. Our results suggest that, if there is a difference in the distribution within the topology of the cosmic web between LSBGs and HSBGs, it does not seem to be significantly strong, reaching a maximum of 20\% in the most appreciable case. Therefore, we can conclude that the environment on a large scale is not as important as group membership and, of course, the galaxy's own halo in the determination of the LSBG nature of galaxies.

\section{Discussion }
\label{sec:Conclusions}

In the present work, we employed a sample of simulated galaxies drawn from the IllustrisTNG project, as a follow-up of \citet{PerezMontano22}, which allowed us to characterize the environment of LSBGs at scales comparable with the size of their host halo, up to their distribution along the cosmic web of the large-scale structure. Our sample includes galaxies with stellar masses in the range $M_{*}$ $> 10^{9}$ M$_\odot$, where LSBGs are classified according to their $r-$band surface brightness within the projected effective radius $r_{50}$. We adopted a kinematic criterion (\citealt{Sales10,Sales12}) to characterize the morphology of the galaxies in the sample, and employed some supplementary data to characterize the overdensity field \citep{Vogelsberger14} and the large-scale spatial distribution \citep{Duckworth20a,Duckworth20b}.

Motivated by a recent review by \citet{McGaugh21}, which emphasizes that the primary distinctions between LSBGs and HSBGs may be attributed to two distinct hypotheses -variations in the spin parameter of the dark matter halo, and variations in halo density- we conducted a comparative analysis of the median values of $\lambda$ and $c_{200}$ in relation to their stellar mass of LSBGs and HSBGs. Our findings reveal that the median value of $\lambda$ for LSBGs consistently surpasses that of HSBGs, irrespective of their morphological type. Meanwhile, the distributions of $c_{200}$ are statistically indistinguishable (Fig. \ref{fig:c200}). This trend clearly supports the hypothesis that variations in the spin is the key parameter determining the low surface brightness nature of these systems, rather than variations in the halo density. Interestingly, this results is in contrast with \citet{Kulier20}, who found that LSBGs are more likely to inhabit more concentrated halos. The authors argue that this is caused by the strong correlation between the age of central galaxies and their star formation rates \citep{Matthee17,Matthee19} found in EAGLE: older galaxies reside within more concentrated halos. Given that there is no difference in the stellar ages of central LSBGs and HSBGs \citep{PerezMontano22}, no differences are expected in the values of $c_{200}$ for galaxies in TNG100. 

The analysis of the shapes of the dark matter halos inhabited by the galaxies in our sample, indicates that LSBGs tend to reside within more spherical halos than for their high surface brightness counterparts, although the distributions of their axis ratios are not dramatically different (Fig. \ref{fig:axis_full}). The absence of galaxies living in prolate halos could be an indication that most of the galaxies in our sample, particularly those corresponding to LSBGs, have not experienced `dry' mergers, as suggested by the recent study of \citet{Li18}, who found that most of the galaxies living within prolate halos have experienced these major `dry' mergers in the past.  This argument, together with the fact that LSBGs are mostly gas-rich systems \citep{PerezMontano22}, reinforces the hypothesis that the formation of an extended component around galaxies could be mainly due to the accretion of material coming from another gas-rich galaxy, in line with previous studies (\citealt{Zhu18, DiCintio19, Zhu23}). Similarly, \citet{Athanassoula13} found that the halo geometry has a strong impact over the baryonic component of the galaxies, such that less spherical halos favour the early formation of stellar bars. This result is in agreement with Fig. \ref{fig:axis_full}, in which the departure from spherical halos will explain the absence of bars in LSBGs \citep{Impey96,Honey16,Cerv17}, which is the case of the simulated galaxies in TNG100 (Chim et al. in prep.).

The existence of an alignment between stellar and dark matter angular momentum vectors has been highlighted by different authors as a natural consequence of the co-evolution of proto-galaxies together with their associated proto-halo, and the dynamical response of the alignment to the assembly of the baryonic disk (e.g., \citealt{Bailin05,Bett10,Velliscig15}). We computed the specific angular momentum of the dark matter component to obtain the angle between $\vec{j}_{*}$ and $\vec{j}_{200}$, finding that the median values of $\theta_{*,\rm DM}$ are systematically lower for LSBGs than in HSBGs, which implies a stronger alignment between both components in LSBGs, favoring a total angular momentum conservation. In a previous work by \citet{Bullock01}, the authors study the angular momentum profiles and the alignment of the angular momentum throughout different ``shells'' within the whole halo volume, of a sample of halos drawn from an N-body simulation. They conclude that halos with misaligned angular momentum distributions are unlikely to host large disk galaxies, such as the ones presented by the LSBGs population.

Extending our analysis to larger scales, we explored the impact of the group environment in which LSBGs reside by quantifying the fraction of satellite LSBGs as a function of the 3D distance $r$ to the center of their corresponding group, normalized by the characteristic main halo size $R_{200}$. We found that the fraction of LSBGs rises with $r$ (Fig. \ref{fig:groups}), indicating that the outer parts of galaxy groups represent a more favorable environment for the formation and/or preservation of LSBGs. For low/medium mass galaxies a clear dependence of $f_{\mathrm{LSB}}$ with $r$ is observed, but not for high-mass galaxies, suggesting that these later are massive enough to ``survive'' violent events. Interestingly, we noted a lack of rotation-dominated LSBGs at $r/R_{200} > 0.4$, compared to dispersion-dominated galaxies which could be an indication of two different evolutionary scenarios: LSBGs with rotation-dominated morphologies are most likely to ``survive'' and remain as LSBGs when they inhabit less-dense environments with no violent interactions that eventually destroy the disks, or transform them into spheroids. This scenario is in good agreement with \citet{Kulier20}, who found that LSBGs are further away from their closest neighbour than HSBGs, implying that galaxies are not only transformed into LSBGs within isolated environments, but also in denser environments, where massive galaxies accrete their neighbours. These close encounters are able to disrupt galaxies leading to the formation of an extended stellar component around them, and the transition from disks to spheroids caused by ram pressure stripping and tidal interactions could explain the lower number of rotation-dominated LSBGs found within galaxy groups \citep{Koda15}.

The relative isolation in which LSBGs reside is extensively pointed out by different authors across the literature (\citealt{Rosenbaum09,Galaz11,PerezMontano19,Tanoglidis21}), however, this does not necessarily imply that LSBGs are formed exclusively within low-density environments. In fact, galaxy mergers and encounters with nearby, low-mass galaxies are also a feasible scenario of LSBGs formation due to the accretion of stellar and gas components, promoting the formation of a faint, extended component around massive galaxies \citep{Kulier20}. We explored the fraction of LSBGs over the $M_{*}$ vs density field $\delta$ plane, finding that, at fixed stellar mass, the fraction of central LSBGs does not depend strongly with the environment, suggesting that the local environmental density is not a key factor for the formation and/or evolution of central LSBGs. 

By contrast, satellite LSBGs are found in a wider variety of environments, with satellite rotation-dominated (dispersion-dominated) populating the low (high) density regime. Mechanisms such as the tidal stripping of the stellar component are able to drastically reduce the stellar mass of LSBGs, while keeping their sizes constant, transforming galaxies with a strong tidal stripping into galaxies with a lower surface brightness \citep{Sales20}. Therefore, it is plausible that LSBGs may have been formed in ``average'' or even high density environments (as proposed by \citealt{Martin19}), but eventually LSBGs are found to be isolated due to the fact that their close neighbours have already been ripped out, as pointed out by \citet{Kulier20}. The absence of rotation-dominated LSBGs at high densities reinforces the idea that these could be completely destroyed, accreted and/or even transformed into a dispersion-dominated due to interactions.

This latter scenario was proposed by \citet{Sales20} to explain the origin of UDGs that were formed within galaxy clusters. These authors employed a sample of UDGs drawn from TNG100, so we can expect that a similar mechanism is occurring in our LSBGs sample. Moreover, these mechanisms would also explain the formation of LSBGs in galaxy groups with the variety of sizes and morphologies found in our sample. Taken together, our results indicate that the density field has a different effect depending on the morphology of the galaxies, such that rotation-dominated LSBGs have been less exposed to strong interactions than dispersion-dominated LSBGs. This implies that the evolution of dispersion-dominated LSBGs is intimately linked with high-density environments, where mergers, tidal interactions and ram-pressure stripping phenomena are responsible for their nature. Our results are in good agreement with what was found by \citet{Saburova21} and \citet{Zhu23}, who found that the lack of major mergers between $z=1-3$ is crucial for the survival of large stellar disks.

We also employed the \texttt{Halotools} package \citep{Hearin17} to compute the cross-correlation function of LSBGs and HSBGs with respect to the full TNG100 galaxy population. Previous studies (\citealt{Bothun93,Mo94}) had found that LSBGs are less strongly clustered than HSBGs, a trend we confirmed in our galaxy sample at scales of $r \lesssim 0.4$ Mpc (Fig. \ref{fig:tpcf_full}). Above this limit we note an inverse trend. When segregating our sample by morphological type, we note that the tpcf of rotation-dominated LSBGs has a lower amplitude than of rotation-dominated HSBGs at all scales, while the rise in the amplitude of the tpcf for the full sample is dominated by the effect of early-type LSBGs residing in high-density environments at scales of $r >$ 0.4 Mpc, which is in a good agreement with previous studies \citep{MEinasto91,Mo92,Zehavi11}. This result highlights that rotation-dominated LSBGs may have formed due to secular evolution, and weak interactions, rather than external violent processes, whereas dispersion-dominated LSBGs are more likely to be formed via external processes, such as mergers, ram-pressure stripping or tidal torques. 

The clustering on scales smaller than 1$h^{-1}$Mpc, which is known as ``the one-halo term'' \citep{PeacockSmith00,Benson00}, is caused by pairs of galaxies within the same parent halo. On the other hand, the clustering on scales larger than a typical halo ($\sim 1-2 h^{-1}$Mpc) corresponds to pairs of galaxies occupying separate halos, also known as the ``two-halo term''. Our results indicate that LSBGs are less clustered in scales $<$ 0.4 Mpc, suggesting that LSBGs do not have close neighbours within their virial radius, and highlighting that processes of galaxy formation and evolution depend strongly on the environment, at least on scales within 1-2 Mpc, which are comparable with the mean size of galaxy clusters \citep{MEinasto91}. The high clustering found in dispersion-dominated LSBGs is a good reflection of the high-density environments in which these galaxies reside, at least in scales above $\sim$ 200 kpc.

Finally, we characterized the distribution of our galaxy sample within the large-scale structure by analyzing the distance to the nearest cosmic web structure, namely voids, walls, filaments, nodes, and saddle points. We divided our galaxy sample between central/satellite galaxies, and considered rotation-dominated/dispersion-dominated morphologies. We did not find significant differences in the large-scale environment between LSBGs and HSBGs in many of them, but for some counted exceptions (Fig. \ref{fig:cumulative}). For central galaxies, we found that dispersion-dominated LSBGs are in general closer to nodes than HSBGs, implying that most of the central dispersion-dominated galaxies are closer to the center of their residing galaxy cluster, and making them subject to frequent galaxy interactions which are responsible of their nature. By contrast, no differences are found in the large-scale distribution between central rotation-dominated LSBGs and HSBGs, which is in good agreement with previous observational studies (e.g., \citealt{Mo94,RomanTrujillo17,PerezMontano19}). Our result highlights the importance of the environment in the morphology of LSBGs, due to the fact that dispersion-dominated LSBGs are found in denser environments (such as nodes), implying that interactions are the main responsible for their nature, in contrast with rotation-dominated LSBGs, whose excess of interactions could even destroy them.

In the case of satellite galaxies, LSBGs are found to be further away from their closest filament than HSBGs with the same morphology, in good agreement with previous observational studies \citep{Rosenbaum09,PerezMontano19}. These authors argued that galaxies formed in low density environments (such as LSBGs) can migrate towards the outer regions of filaments due to gravitational infall. Therefore, LSBGs may have formed within void regions, but most of them have migrated to the edges of the filaments. The very small differences found in the large-scale structure distribution between LSBGs and HSBGs (about 20\% in the most representative case) allow us to affirm that the large-scale structure of the Universe has a minor role in the determination of the low surface brightness nature of the galaxies in our sample. 

This later result is in line with \citet{Goh19}, where the authors study the effects of the local environment and the cosmic web over some key properties of DM halos such as the spin parameter, the concentration index and the halo geometry, finding no significant variation among them across different large-scale environments. However, such halo properties do vary as a function of the environmental densities. \citet{Kuutma20} found that the filamentary structure has a very limited effect on the properties of galaxies, depending mostly on their local environment density field. In \citet{Das22}, the authors found that galaxy pairs with a projected separation $<$ 50 kpc are more star-forming in filaments than pairs hosted in walls. Similarly, at smaller pair separations, galaxy pairs within wall-like structures are significantly redder than galaxy pairs in filaments. Even when large-scale structures have a fundamental role in galaxy interactions that affect some properties of galaxies, such as their SFRs and color, they will equally affect LSBGs and HSBGs. Therefore, we can conclude that the density field, rather than the large-scale distribution should be the main factor affecting halo properties.

\section{Conclusions}
Altogether, the main results of the current work can be summarized as follows:

(i)  At fixed stellar mass, LSBGs exhibit systematically higher halo spin parameters than HSBGs. This discrepancy in the distribution of the spin parameter ($\lambda$) is likely the primary factor underlying their distinctive nature, rather than variations in the density concentration of the DM halo.

(ii) In LSBGs, the specific stellar angular momentum ($j_{*}$) displays a more pronounced alignment with the dark matter component ($j_{200}$) and exhibits a narrower distribution than the one observed in HSBGs.

(iii) While the majority of galaxies in our sample are predominantly immersed within spherical DM halos, LSBGs tend to occupy slightly more spherical halos compared to HSBGs.

(iv) The study of the clustercentric distance, the overdensity field, and the cross-correlation function found for LSBGs and HSBGs of different morphologies reflects that rotation-dominated LSBGs are most likely to be found within low-density environments, which enables their persistence as LSBGs throughout cosmic time. In contrast, dispersion-dominated LSBGs (mostly with $M_{*} < 10^{10} \Msun$) may have formed in high-density environments, where vigorous interactions and mergers might trigger their transformation into this kind of systems.

(vi) When studying the influence of the large-scale structure on their observed properties, there is generally no significant disparity in the distribution within the cosmic web between LSBGs and HSBGs, barring specific exceptions such as nodes and filaments. 

In summary, our findings align well with the hypothesis suggesting that the formation mechanism of LSBGs is predominantly influenced by the evolution of angular momentum and spin parameters rather than variations in halo density. The local environment, particularly within galaxy groups, emerges as a critical factor shaping these phenomena, impacting the evolution \citep{MEinasto91}. Given the observed differences in galactic environments among various galaxy types, a comprehensive investigation into the evolution of surface brightness and morphology becomes desirable. Understanding key factors, such as the mass accretion history and the frequency and alignment of mergers, might result revealing. Examining the evolutionary trajectories can shed light on whether LSBGs have consistently maintained their nature or have undergone recent transitions from HSBGs to LSBGs, possibly undergoing one or multiple transformations in surface brightness over cosmic time. 

If such transitions indeed occur, it becomes crucial to ascertain whether they stem from gradual internal changes (secular evolution) or are triggered by external influences. Notably, beyond surface brightness and morphology (e.g., \citealt{Abraham01,DiCintio19,Park22}), various other properties undergo changes throughout a galaxy's lifetime, including accretion histories, the growth of supermassive black holes within them, and their interconnectedness with the evolution of angular momentum and spin parameters \citep{RodGom22,Singh23}.

\section*{Acknowledgements}

The authors thank Gaspar Galaz, Ricardo Chávez-Murillo, Miguel Aragón-Calvo, Simon Kemp and Rosa Amelia González-Lópezlira for their useful comments and discussions. We also thank the anonymous referee for the thorough reading of the original manuscript and for the insightful report that
helped to improve the quality of the paper and the clarity of the results presented. Luis Enrique Pérez-Montaño and Bernardo Cervantes Sodi also acknowledge the financial support provided by PAPIIT project IA103520 and IN108323 from DGAPA-UNAM. Luis Enrique Pérez-Montaño and Go Ogiya were supported by the National Key Research and Development Program of China (Grant No. 2022YFA1602903), the National Natural Science Foundation of China (Grant No. 12373004) and the Fundamental Research Fund for Chinese Central Universities (Grant No. NZ2020021, No. 226-2022-00216).

\section*{Data availability}

The data from the IllustrisTNG simulations employed in this work is publicly available at the website \href{https://www.tng-project.org}{https://www.tng-project.org} \citep{Nelson19}.



\bsp	
\label{lastpage}
\end{document}